\input{config/fixes}

\documentclass[a4paper,11pt]{article}

\usepackage{jcappub}
\usepackage[UKenglish]{babel}
\usepackage{bbm}
\usepackage[numbered]{bookmark}
\usepackage[inline]{enumitem}
\usepackage{etoolbox}
\usepackage{fontawesome}
\usepackage[T1]{fontenc}
\usepackage{journalref}
\usepackage{lmodern}
\usepackage{mathrsfs}
\usepackage{mathtools}
\usepackage{microtype}
\usepackage{physics}
\usepackage{ragged2e}
\usepackage{subcaption}
\usepackage{siunitx}
\usepackage{textgreek}
\usepackage{upgreek}
\usepackage{xparse}
\usepackage{cleveref}
\usepackage[usenames,dvipsnames]{xcolor}

\DeclareFontFamily{U}{mathx}{\hyphenchar\font45}
\DeclareFontShape{U}{mathx}{m}{n}{
    <5> <6> <7> <8> <9> <10>
    <10.95> <12> <14.4> <17.28> <20.74> <24.88>
    mathx10
}{}
\DeclareSymbolFont{mathx}{U}{mathx}{m}{n}

\DeclareMathAccent{\widebar}{0}{mathx}{"73}
\DeclareMathAccent{\widehat}{0}{mathx}{"70}
\DeclareMathAccent{\widetilde}{0}{mathx}{"72}

\appto\bibfont{\RaggedRight}

\preto\subequations{\ifhmode\unskip\fi}

\crefname{figure}{figure}{figures}

\setlist[enumerate]{label=\arabic*)}

\graphicspath{{graphics/}}

\DeclareSIUnit{\h}{\text{$h$}}            
\DeclareSIUnit{\parsec}{pc}               
\DeclareSIUnit{\stdsig}{\text{$\sigma$}}  

\ExplSyntaxOn
\DeclareExpandableDocumentCommand{\IfNoValueOrEmptyTF}{mmm}{
    \IfNoValueTF{#1}{#2}{{\tl_if_empty:nTF {#1} {#2} {#3}}}
}
\ExplSyntaxOff

\makeatletter
\DeclareDocumentCommand{\@arguments}{
    t\big t\Big t\bigg t\Bigg g o d() d||
}{
    \IfBooleanTF{#1}{%
        \let\ltag\bigl \let\rtag\bigr
    }{%
        \IfBooleanTF{#2}{%
            \let\ltag\Bigl \let\rtag\Bigr
        }{%
            \IfBooleanTF{#3}{%
                \let\ltag\biggl \let\rtag\biggr
            }{%
                \IfBooleanTF{#4}{%
                    \let\ltag\Biggl \let\rtag\Biggr
                }{%
                    \let\ltag\left \let\rtag\right
                }%
            }%
        }%
    }%
    \IfNoValueTF{#5}{%
        \IfNoValueTF{#6}{%
            \IfNoValueTF{#7}{%
                \IfNoValueTF{#8}{%
                    ()
                }{%
                    \ltag\lvert{#8}\rtag\rvert
                }%
            }{%
                \ltag(#7\rtag) \IfNoValueTF{#8}{}{|#8|}
            }%
        }{%
            \ltag[#6\rtag]
            \IfNoValueTF{#7}{}{(#7)}
            \IfNoValueTF{#8}{}{|#8|}
        }%
    }{%
        \ltag\lbrace#5\rtag\rbrace
        \IfNoValueTF{#6}{}{[#6]}
        \IfNoValueTF{#7}{}{(#7)}
        \IfNoValueTF{#8}{}{|#8|}
    }%
    \ifnum\z@=`{\fi}
}
\makeatother

\DeclareDocumentCommand{\argopen}{s}{%
    \IfBooleanTF{#1}{%
        \mathopen{}\mathclose\bgroup%
    }{%
        \mathopen{}\mathclose\bgroup\left%
    }
}
\DeclareDocumentCommand{\argclose}{s}{%
    \IfBooleanTF{#1}{\egroup}{\aftergroup\egroup\right}%
}

\makeatletter
\let\originalleft\left
\renewcommand{\left}{\mathopen{}\mathclose\bgroup\originalleft}
\let\originalright\right
\renewcommand{\right}{\aftergroup\egroup\originalright}
\DeclareDocumentCommand{\oper}{s m}{%
    \IfBooleanTF{#1}{%
        \operatorname{#2}%
    }{%
        \operatorname{#2}\!{\ifnum\z@=`}\fi\@arguments%
    }%
}
\DeclareDocumentCommand{\func}{s m}{%
    \IfBooleanTF{#1}{%
        #2%
    }{%
        \mathop{}\!#2{\ifnum\z@=`}\fi\@arguments%
    }%
}
\makeatother

\ExplSyntaxOn
\NewDocumentCommand{\burl}{m}{
    \texttt{\break_insert:n{#1}}
}  
\NewDocumentCommand{\bhref}{m}{
    \href{\break_replace:n{#1}}{\texttt{\break_insert:n{#1}}}
}  

\cs_new:Nn \break_replace:n {
    \tl_set:Nn \l_tmpa_tl { #1 }
    \tl_replace_all:Nnn \l_tmpa_tl { | } { }
    \l_tmpa_tl
}
\cs_new_protected:Nn \break_insert:n {
    \seq_set_split:Nnn \l_tmpa_seq { | } { #1 }
    \seq_use:Nn \l_tmpa_seq { \allowbreak }
}
\ExplSyntaxOff

\newdimen\capht
\newdimen\desdp
\newcommand{\orcid}[1]{%
    \settoheight{\capht}{\uppercase{X}}%
    \settodepth{\desdp}{g}%
    \hspace*{0.5\capht}%
    \raisebox{-0.4\desdp}{%
        \href{https://orcid.org/#1}{%
            \includegraphics[height=1.05\capht]{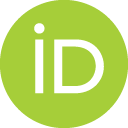}%
        }%
    }%
}
\newcommand{\codename}[1]{\textsc{#1}}  
\newcommand{\is}{is}                    

\newcommand{\mat}[1]{\boldsymbol{\mathsf{#1}}}  

\NewDocumentCommand{\ifrac}{o m m}{%
    \IfNoValueTF{#1}{{%
        \argopen.\mkern-2mu{#2}\mkern2mu%
        \middle/%
        \mkern2mu{#3}\mkern-2mu\argclose.%
    }}{#2\,#1|\,#3}%
}  

\makeatletter

\renewcommand{\geq}{\geqslant}
\makeatother   
\makeatletter

\renewcommand{\leq}{\leqslant}
\makeatother   

\newcommand{\defeq}{=}

\makeatletter
\newcommand{\transT}{{\mathpalette\@transT{}}}
\newcommand{\@transT}[2]{\raisebox{\depth}{$\m@th#1\intercal$}}
\makeatother
\newcommand{\trs}[1]{{#1}^\transT}   

\newcommand{\conj}[1]{{#1}^\ast}       

\newcommand{\estm}[1]{\widehat{#1}}  
\newcommand{\wind}[1]{\tilde{#1}}

\newcommand{\comma}{\,,}
\newcommand{\fstop}{\,.}
\newcommand{\mquad}[1][1]{\hspace*{#1em}\ignorespaces}

\newcommand{\lunit}{\per\h\mega\parsec}
\newcommand{\wunit}{\h\per\mega\parsec}

\newcommand{\e}{\mathrm{e}}   
\newcommand{\im}{\mathrm{i}}  

\newcommand{\dirac}{\delta^{(\mathrm{D})}}

\newcommand{\shY}[2]{Y_{#1}^{#2}}

\newcommand{\shy}[2]{y_{#1}^{#2}}
\newcommand{\shyc}[2]{\conj{y_{#1}^{#2}}}
\newcommand{\wigthreej}[6]{\mqty(#1 & #2 & #3 \\ #4 & #5 & #6)}

\newcommand{\hwig}[3]{\wigthreej{#1}{#2}{#3}{0}{0}{0}}

\newcommand{\wigninej}[9]{\qty{\mqty{#1 & #2 & #3 \\ #4 & #5 & #6 \\ #7 & #8 & #9}}}

\newcommand{\nbar}{\bar{n}}
\newcommand{\dn}{\delta{n}}
\newcommand{\oden}{\delta}
\newcommand{\zetabar}{\bar{\zeta}}

\newcommand{\los}{\vu{n}}
\newcommand{\vx}{\vb{x}}
\newcommand{\vr}{\vb{r}}
\newcommand{\vk}{\vb{k}}

\newcommand{\vux}{\vu{x}}
\newcommand{\vur}{\vu{r}}
\newcommand{\vuk}{\vu{k}}

\newcommand{\rf}{\mathrm{ref}}
\newcommand{\sys}{\mathrm{s}}

\newcommand{\mW}{\mat{W}}

\newcommand{\vB}{\vb{B}}
\newcommand{\vwB}{\wind{\vB}}
\newcommand{\vwu}{\wind{\vb{u}}}

\title{Window convolution of the galaxy clustering bispectrum}

\author[a]{M.~S.~Wang\orcid{0000-0002-2652-4043},}
\author[a]{F.~Beutler\orcid{0000-0003-0467-5438},}

\affiliation[a]{%
    Institute for Astronomy, University of Edinburgh, \\
    Royal Observatory Edinburgh, Blackford Hill, Edinburgh, EH9 3HJ, UK
}

\author[b]{J.~Aguilar,}
\author[c]{S.~Ahlen\orcid{0000-0001-6098-7247},}
\author[d]{D.~Bianchi\orcid{0000-0001-9712-0006},}
\author[e]{D.~Brooks,}
\author[b]{T.~Claybaugh,}
\author[f]{A.~de~la~Macorra\orcid{0000-0002-1769-1640},}
\author[e]{P.~Doel,}
\author[g,e]{A.~Font-Ribera\orcid{0000-0002-3033-7312},}
\author[h,i,j]{E.~Gazta\~naga,}
\author[k]{G.~Gutierrez,}
\author[l,m]{K.~Honscheid\orcid{0000-0002-6550-2023},}
\author[n]{C.~Howlett\orcid{0000-0002-1081-9410},}
\author[o]{D.~Kirkby\orcid{0000-0002-8828-5463},}
\author[b]{A.~Lambert,}
\author[b]{M.~Landriau\orcid{0000-0003-1838-8528},}
\author[p,g]{R.~Miquel,}
\author[q,r]{G.~Niz\orcid{0000-0002-1544-8946},}
\author[s]{F.~Prada\orcid{0000-0001-7145-8674},}
\author[t]{I.~P\'erez-R\`afols\orcid{0000-0001-6979-0125},}
\author[u]{G.~Rossi,}
\author[v]{E.~Sanchez\orcid{0000-0002-9646-8198},}
\author[b]{D.~Schlegel,}
\author[w]{M.~Schubnell,}
\author[x]{D.~Sprayberry,}
\author[w]{G.~Tarl\'e\orcid{0000-0003-1704-0781}}
\author[x]{and B.~A.~Weaver}

\affiliation[b]{%
    Lawrence Berkeley National Laboratory, \\
    1~Cyclotron Road, Berkeley, CA~94720, USA
}
\affiliation[c]{%
    Department of Physics, Boston University, \\
    590~Commonwealth Avenue, Boston, MA~02215, USA
}
\affiliation[d]{%
    Dipartimento di Fisica ``Aldo Pontremoli'', Universit\`a degli Studi di Milano, \\
    Via Celoria~16, I-20133~Milano, Italy
}
\affiliation[e]{%
    Department of Physics and Astronomy, University College London, \\
    Gower Street, London, WC1E~6BT, UK
}
\affiliation[f]{%
    Instituto de F\'{\i}sica, Universidad Nacional Aut\'onoma de M\'exico, \\
    Circuito de la Investigaci\'on Cient\'{\i}fica, Ciudad Universitaria, \\
    Ciudad de M\'exico, C.P.~04510, M\'exico
}
\affiliation[g]{%
    Institut de F\'isica d'Altes Energies~(IFAE), The Barcelona Institute of Science and Technology, \\
    Campus UAB, 08193~Bellaterra Barcelona, Spain
}
\affiliation[h]{
    Institut d'Estudis Espacials de Catalunya~(IEEC), \\
    Gran Capit\`an~2--4, 08034~Barcelona, Spain
}
\affiliation[i]{
    Institute of Cosmology and Gravitation, University of Portsmouth, \\
    Dennis Sciama Building, Portsmouth, PO1~3FX, UK
}
\affiliation[j]{
    Institute of Space Sciences (ICE-CSIC), \\
    Campus UAB, Carrer de Can Magrans s/n, 08913~Bellaterra Barcelona, Spain
}
\affiliation[k]{%
    Fermi National Accelerator Laboratory, \\
    PO~Box~500, Batavia, IL~60510, USA
}
\affiliation[l]{%
    Center for Cosmology and AstroParticle Physics, The Ohio State University, \\
    191~West Woodruff Avenue, Columbus, OH~43210, USA
}
\affiliation[m]{%
    Department of Physics, The Ohio State University, \\
    191~West Woodruff Avenue, Columbus, OH~43210, USA
}
\affiliation[n]{
    School of Mathematics and Physics, University of Queensland, \\
    St Lucia, Brisbane, QLD~4072, Australia
}
\affiliation[o]{
    Department of Physics and Astronomy, University of California, Irvine, \\
    Frederick Reines Hall, Irvine, CA~92697, USA
}
\affiliation[p]{%
    Instituci\'o Catalana de Recerca i Estudis Avan\c{c}ats, \\
    Passeig de Llu\'{\i}s Companys~23, 08010~Barcelona, Spain
}
\affiliation[q]{
    Departamento de F\'{\i}sica, Universidad de Guanajuato, \\
    DCI-Campus Le\'{o}n, Loma del Bosque~103, Le\'{o}n, Guanajuato, C.~P.~37150, M\'{e}xico
}
\affiliation[r]{
    Instituto Avanzado de Cosmolog\'{\i}a A.~C., \\
    San Marcos 11--Atenas 202, Magdalena Contreras, Ciudad de M\'{e}xico, C.~P.~10720, M\'{e}xico 
}
\affiliation[s]{%
    Instituto de Astrof\'isica de Andaluc\'ia (CSIC), \\
    Glorieta de la Astronom\'ia~s/n, E-18008~Granada, Spain
}
\affiliation[t]{%
    Departament de F\'isica, EEBE, Universitat Polit\`ecnica de Catalunya, \\
    Eduard Maristany~10, 08930~Barcelona, Spain
}
\affiliation[u]{%
    Department of Physics and Astronomy, Sejong University, \\
    Gwangjin-Gu, Seoul, 143-747, Korea
}
\affiliation[v]{%
    Centro de Investigaciones Energ\'eticas, Medioambientales y Tecnol\'ogicas (CIEMAT), \\
    Avenida Complutense~40, E-28040~Madrid, Spain
}
\affiliation[w]{%
    Department of Physics, University of Michigan, \\
    Ann Arbor, MI~48109, USA
}
\affiliation[x]{%
    NSF NOIRLab, \\
    950~N.~Cherry~Ave., Tucson, AZ~85719, USA
}

\emailAdd{mikeshengbo.wang@ed.ac.uk}
\emailAdd{florian.beutler@ed.ac.uk}

\abstract{%
In galaxy survey analysis, the observed clustering statistics do not directly match theoretical predictions but rather have been processed by a window function that arises from the survey geometry including the sky footprint, redshift-dependent background number density and systematic weights.
While window convolution of the power spectrum is well studied, for the bispectrum with a larger number of degrees of freedom, it poses a significant numerical and computational challenge.
In this work, we consider the effect of the survey window in the tripolar spherical harmonic decomposition of the bispectrum and lay down a formal procedure for their convolution via a series expansion of configuration-space three-point correlation functions, which was first proposed by Sugiyama~et~al.~(2019).
We then provide a linear algebra formulation of the full window convolution, where an unwindowed bispectrum model vector can be directly premultiplied by a window matrix specific to each survey geometry.
To validate the pipeline, we focus on the Dark Energy Spectroscopic Instrument~(DESI) Data Release~1~(DR1) luminous red galaxy~(LRG) sample in the South Galactic Cap~(SGC) in the redshift bin \(\num{0.4} \leq z \leq \num{0.6}\).
We first perform convergence checks on the measurement of the window function from discrete random catalogues,
and then investigate the convergence of the window convolution series expansion truncated at a finite of number of terms as well as the performance of the window matrix.
This work highlights the differences in window convolution between the power spectrum and bispectrum, and provides a streamlined pipeline for the latter for current surveys such as DESI and the \textit{Euclid} mission.

}


\arxivnumber{2411.14947}

\makeatletter
\hypersetup{
    pdftitle={\@title},
    pdfauthor={Mike (Shengbo) Wang, Florian Beutler},
    pdfkeywords={
        galaxy clustering,
        redshift surveys
    }
}
\makeatother

\begin{document}

\maketitle

\flushbottom

\section{Introduction}
\label{sec:introduction}

Measurements of cosmological observables usually cannot be directly compared to their theoretical predictions not only because they are noisy real{\is}ations, but also because they have been processed by a window function that filters the underlying signal.
In galaxy clustering analysis, the window function encodes the survey geometry and includes selection functions of the background number density, as well as any weights applied to compensate for systematic effects or to enhance the signal-to-noise ratio~\citep{Feldman:1994}.
For correlation functions in configuration space, the window functions are locally multiplicative in nature and thus relatively easy to account for; for instance, the Landy--Szalay estimator of correlation functions largely removes the effect of the window function through multiplet counting with the random catalogue~\cite{Landy:1993}.
In Fourier space, the window function instead convolves with the polyspectra such as the power spectrum and bispectrum.
This mixes clustering modes on different scales, and modulates both the shape and the amplitude of the clustering statistics~\citep{Peacock:1991}; moreover, since the window function itself is typically anisotropic, it couples to anisotropies arising from redshift-space distortions~(RSD) and the Alcock--Paczy\'nski~(AP) effect~\citep{Hamilton:1998,Beutler:2014}, and therefore must be accounted for in their statistical analysis.

One way to disentangle the window function from the power spectrum and bispectrum is to deconvolve it using the quadratic and cubic (near-)optimal estimators~\citep[e.g.][]{Tegmark:1997a,Tegmark:1997b,Hamilton:1997a,Hamilton:1997b,Tegmark:1998,Philcox:2021a,Philcox:2021b}, but these are numerically difficult to obtain and require the inversion and differentiation of covariance and Fisher matrices at the map pixel level.
Fundamentally, deconvolution is an inverse operation (akin to division in configuration space), and so it may suffer from ill-conditioning especially with noisy estimates of clustering statistics.

The alternative approach, which has become standard for the analysis of the galaxy clustering power spectrum, is to forward model the window function by performing convolution as multiplication in configuration space~\citep{Wilson:2016,Beutler:2016}.
This is particularly advantageous for compressed clustering statistics such as Legendre multipoles of the correlation function or the power spectrum (\(\xi_L\) or \(P_L\)), where the window function multipoles~\(Q_L\) can be measured in a similar way, and multipoles in configuration and Fourier spaces are related through the Hankel transform.
In addition, for predefined binning, the entire procedure is a combination of linear algebra operations and can be cast as matrix multiplications for fast computation~\cite{Blake:2018,Beutler:2021}.

Although the treatment of the window function for power spectrum has been standard{\is}ed this way in recent analyses of the 6-degree Field Galaxy Redshift Survey~(6dFGRS), the Baryon Oscillation Spectroscopic Survey~(BOSS) and its extension eBOSS~\citep[e.g.][]{Beutler:2016,Beutler:2017,Blake:2018,Gil-Marin:2020,deMattia:2021,Neveux:2020}, it remains challenging for three-point clustering statistics in Fourier space, where perturbative models of the bispectrum such as the effective field theory of large-scale structure are presented~\citep[e.g.][]{Baumann:2012,Carrasco:2012,Hertzberg:2014,Senatore:2015a,Senatore:2015b,dAmico:2020,Ivanov:2020}.
This is because higher-order clustering statistics have many more degrees of freedom; for instance, whereas the power spectrum is a function of a single wavevector~\(\vk\) under the assumption of statistical homogeneity, the bispectrum is a function of two wavevectors defining a closed triangle also under statistical homogeneity.

In the analysis of BOSS galaxy catalogues, refs.~\citep{Gil-Marin:2015,Gil-Marin:2016} have replaced the bispectrum window convolution with convolution of the two-point window function with each power spectrum kernel in the bispectrum model; this is a mathematically inequivalent procedure and an inaccurate approximation on large scales.
More recently, ref.~\citep{Alkhanishvili:2023} has suggested the deployment of deep neural networks to `learn' the effect of the window function on the bispectrum from a large suite of training simulations; aside from the computational cost of this approach which needs to be repeated for each window function, this proof-of-concept study has had to assume a relation between the unwindowed and windowed bispectra that is a local and isotropic response function in Fourier space.
For the commonly considered Scoccimarro bispectrum multipoles~\(\func{B_L^M}(k_1, k_2, k_3)\) in the basis proposed by ref.~\citep{Scoccimarro:2015}, ref.~\citep{Pardede:2022} has laid down the analytical expressions for the window convolution, but the window function is decomposed into a much larger set of multipole components with five additional spherical degrees~\(\ell\), each requiring the summation over multiple spherical orders~\(m\).
This induces a significant computational cost, even when they can be recast in a form amenable to the fast Fourier transform~(FFT).

In this work, we consider an alternative decomposition of three-point clustering statistics in the tripolar spherical harmonic~(TripoSH) basis, which was first proposed by ref.~\citep{Sugiyama:2019}.
In this decomposition, the third internal angle (or equivalently, its corresponding side) of the bispectrum triangles is integrated out, resulting in multipoles~\(\func{B_{\ell_1 \ell_2 L}}(k_1, k_2)\) with only two wavenumbers and two additional spherical degrees~\(\ell_1, \ell_2\).%
\footnote{Throughout this work, we consider only single-digit spherical degrees~\(\ell_{1,2} < 10\) and do not delimit the subscripts with commas.}
In contrast to the Scoccimarro multipoles which have three wavenumbers, this is a form of data dimensionality reduction, as the bispectrum multipoles are now effectively two-dimensional rather than three-dimensional quantities, which reduces the computational cost, especially for covariance matrix estimation.
The three-point correlation function~(3PCF) admits a similar expansion into multipoles that are related to the bispectrum multipoles through a double spherical Bessel transform.
More crucially, as demonstrated in ref.~\citep{Sugiyama:2019} and discussed later in this work, the window function in the TripoSH formulation can be decomposed into similar multipoles that can be measured in the same way as for the bispectrum and 3PCF multipoles using FFTs.
The full window convolution procedure is then analogous to that of two-point clustering statistics as laid down by refs.~\cite{Wilson:2016,Beutler:2016}, where in configuration space the windowed 3PCF multipoles are series expansions in terms of the window function multipoles and the unwindowed 3PCF multipoles.\footnote{Note that unlike the Landy--Szalay type of estimators which use multiplet counting with the random catalogue to remove the window function, these FFT-based correlation function estimators still include the window effect.}
Whereas ref.~\citep{Sugiyama:2019} only considers the leading-order term in the series and ref.~\citep{Sugiyama:2020} focuses on the configuration space, we investigate the importance of non-leading-order terms here in Fourier space.
Moreover, this approach also allows us to propose a simple linear algebra formulation where a given unwindowed bispectrum model vector is premultiplied by a window matrix, the result of which can then be compared with the measured data vector as for the power spectrum~\cite{Beutler:2021}.

This article proceeds as follows.
We first introduce the TripoSH decomposition and write down the FFT-based estimators proposed by ref.~\cite{Sugiyama:2019} in \cref{sec:statistics}.
We then lay out the window convolution procedure in \cref{sec:convolution}, including the numerical implementation of the double spherical Bessel transform required to convert between Fourier- and configuration-space statistics and the linear algebra formulation of a window matrix.
In a series of validation tests detailed in \cref{sec:validation}, we discuss the convergence of window function measurements with respect to the sampling of the number density field, as well as truncation and approximations adopted in the window convolution formula.
We will highlight some technical differences in window convolution between three-point and two-point clustering statistics as well as between Fourier and configuration space, before drawing a conclusion in \cref{sec:conclusion}.

Throughout this work, we will use our publicly available code~\codename{Triumvirate}\footnote{\label{fn:Triumvirate}\href{https://github.com/MikeSWang/Triumvirate}{\texttt{github.{\allowbreak}com/{\allowbreak}MikeSWang/{\allowbreak}Triumvirate}}} to measure all three-point clustering statistics including the window function and to perform window convolution \cite{Wang:2023a}.
We also adopt the following convention for the forward and backward/inverse Fourier transforms (of a three-dimensional field~\(f\)):
\begin{align}
    \func{f}(\vk) = \int \dd[3]{\vx} \e^{-\im\vk\vdot\vx} \func{f}(\vx) \comma \quad \func{f}(\vx) = \int \frac{\dd[3]{\vk}}{(2\uppi)^3} \e^{\im\vk\vdot\vx} \func{f}(\vk) \fstop
\end{align}

\section{Tripolar spherical harmonic decomposition}
\label{sec:statistics}

Given a catalogue of discrete particles, one could assign their positions~\(\vx\) to a regular mesh grid of cells to compute fluctuations in the number density field~\(n(\vx)\) with any weights~\(\func{w}(\vx)\) (such as the Feldman--Kaiser--Peacock scheme~\cite{Feldman:1994}), 
\begin{equation}
    \func{\dn}(\vx) = \func{w}(\vx) \qty[\func{n}(\vx) - \func{\nbar}(\vx)] \fstop
    \label{eq:fluctuation-field}
\end{equation}
Here the background number density~\(\func{\nbar}(\vx)\) is constant for a simulation snapshot, and for a galaxy survey catalogue, it may include the selection function and any additional systematic weights to correct for observational effects such as incompleteness.
In the latter case, \(\func{\nbar}(\vx)\) is typically sampled by a much denser random catalogue but rescaled to match the number density of the data catalogue.

In redshift space where the radial distance to each galaxy is converted from its observed redshift subject to peculiar velocity, clustering statistics such as the 3PCF depends on the line of sight, \(\los\), in addition to the separation~\(\vr_{1,2}\),
\begin{equation}
    \zeta(\vr_1, \vr_2, \los) = \expval{\func{\oden}(\vx + \vr_1) \func{\oden}(\vx + \vr_2) \func{\oden}(\vx)} \comma
    \label{eq:3PCF-local-pp}
\end{equation}
where \(\expval{\cdot}\) denotes the ensemble average, and \(\func{\oden}(\vx) = \func{\dn}(\vx)/\func{\nbar}(\vx)\) is the over-density field.
In the expression above, we have assumed the end-point definition of the line-of-sight vector~\(\los = \vx / \abs{\vx}\) in the local plane-parallel picture.

Analogously, the bispectrum~\(\func{B}(\vk_1, \vk_2, \los)\) in Fourier space is a function of two wave\-vectors~\(\vk_1\), \(\vk_2\) and the line-of-sight vector~\(\los\) in the local plane-parallel picture~\cite{Yamamoto:2006,Scoccimarro:2015}.
Under the assumption of parity symmetry and rotational invariance of the underlying physical processes, one could construct a basis of functions from three spherical harmonics~\citep{Sugiyama:2019},
\begin{equation}
    \func{S_{\ell_1 \ell_2 L}}\big(\vuk_1, \vuk_2, \los) \defeq H_{\ell_1 \ell_2 L}^{-1} \sum_{m_1 m_2 M} \wigthreej{\ell_1}{\ell_2}{L}{m_1}{m_2}{M} \func{\shy{\ell_1}{m_1}}\big(\vuk_1) \func{\shy{\ell_2}{m_2}}\big(\vuk_2) \func{\shy{L}{M}}(\los) \comma
    \label{eq:TripoSH-basis}
\end{equation}
where the Wigner 3-\textit{j} factor
\begin{equation}
    H_{\ell_1 \ell_2 L} \defeq \hwig{\ell_1}{\ell_2}{L}
\end{equation}
enforces \(\qty(\ell_1 + \ell_2 + L) \in 2\mathbb{Z}\) to be even, and the spherical harmonics~\(\shy{\ell}{m} \defeq  \sqrt{\ifrac{4\pi}{(2\ell+1)}} \shY{\ell}{m}\) are normal{\is}ed such that \(\shy{0}{0} \equiv 1\).
In this basis, the bispectrum can be expanded as
\begin{equation}
    \func{B}(\vk_1, \vk_2, \los) = \sum_{\ell_1 + \ell_2 + L \in 2\mathbb{Z}} \func{B_{\ell_1 \ell_2 L}}(k_1, k_2) \func{S_{\ell_1 \ell_2 L}}\big(\vuk_1, \vuk_2, \los) \comma
    \label{eq:TripoSH-expansion}
\end{equation}
where the multipoles are given by
\begin{multline}
    \func{B_{\ell_1 \ell_2 L}}(k_1, k_2) = N_{\ell_1 \ell_2 L} H_{\ell_1 \ell_2 L} \sum_{m_1 m_2 M} \wigthreej{\ell_1}{\ell_2}{L}{m_1}{m_2}{M} \\
    \qquad \times \int \frac{\dd[2]{\vuk_1}}{4\uppi} \frac{\dd[2]{\vuk_2}}{4\uppi} \frac{\dd[2]{\los}}{4\uppi}
    \func{\shyc{\ell_1}{m_1}}\big(\vuk_1) \func{\shyc{\ell_2}{m_2}}\big(\vuk_2) \func{\shyc{L}{M}}(\los) \func{B}(\vk_1, \vk_2, \los)
    \label{eq:bispectrum-multipoles}
\end{multline}
with the prefactor~\(N_{\ell_1 \ell_2 L} \defeq (2\ell_1 + 1) (2\ell_2 + 1) (2L + 1)\).

Similarly, the 3PCF also admits the same expansion as \cref{eq:TripoSH-expansion,eq:TripoSH-basis}, with its multipoles given by
\begin{multline}
    \func{\zeta_{\ell_1 \ell_2 L}}(r_1, r_2) = N_{\ell_1 \ell_2 L} H_{\ell_1 \ell_2 L} \sum_{m_1 m_2 M} \wigthreej{\ell_1}{\ell_2}{L}{m_1}{m_2}{M} \\
    \qquad \times \int \frac{\dd[2]{\vur_1}}{4\uppi} \frac{\dd[2]{\vur_2}}{4\uppi} \frac{\dd[2]{\los}}{4\uppi}
    \func{\shyc{\ell_1}{m_1}}\big(\vur_1) \func{\shyc{\ell_2}{m_2}}\big(\vur_2) \func{\shyc{L}{M}}(\los) \func{\zeta}(\vr_1, \vr_2, \los) \fstop
    \label{eq:3PCF-multipoles}
\end{multline}
These are related to the bispectrum multipoles (eq.~\ref{eq:bispectrum-multipoles}) through a double spherical Bessel transform:%
\footnote{These transforms can be related to and recast as Hankel transforms as commonly known in the literature.}
\begin{subequations}
\label{eq:double-spherical-Bessel-transform}
\begin{align}
    \label{eq:double-spherical-Bessel-transform-forward}
    \func{B_{\ell_1 \ell_2 L}}(k_1, k_2) &= (4\uppi)^2 {\im}^{-\ell_1-\ell_2} \int \dd[2]{r_1} r_1^2 \int \dd[2]{r_2} r_2^2 \func{j_{\ell_1}}(k_1 r_1) \func{j_{\ell_2}}(k_2 r_2) \func{\zeta_{\ell_1 \ell_2 L}}(r_1, r_2) \comma \\
    \label{eq:double-spherical-Bessel-transform-backward}
    \func{\zeta_{\ell_1 \ell_2 L}}(r_1, r_2) &= (4\uppi)^2 {\im}^{\ell_1+\ell_2} \int \frac{\dd{k_1} k_1^2}{(2\uppi)^3} \int \frac{\dd{k_2} k_2^2}{(2\uppi)^3} \func{j_{\ell_1}}(k_1 r_1) \func{j_{\ell_2}}(k_2 r_2) \func{B_{\ell_1 \ell_2 L}}(k_1, k_2) \fstop
\end{align}
\end{subequations}

\subsection{Estimators of three-point clustering statistics}

For convenience, we \emph{define} the spherical harmonic weighted fluctuation field and its Fourier transform,
\begin{equation}
    \func{\dn_\ell^m}(\vx) \defeq \shyc{\ell}{m}(\vux) \dn(\vx) \qand \func{\dn_\ell^m}(\vk) \defeq \int \dd[3]{\vx} \e^{-\im\vk\vdot\vx} \func{\dn_\ell^m}(\vx) \comma
\end{equation}
where the latter can be computed using fast Fourier transform~(FFT) with mesh assignment window corrections applied~\cite{Jing:2005,Sefusatti:2016}.
Assuming the end-point definition of the line of sight, the TripoSH multipoles of the bispectrum and 3PCF in \cref{eq:bispectrum-multipoles,eq:3PCF-multipoles} can be cast in a form compatible with FFT~\cite{Sugiyama:2019}:
\begin{subequations}
\begin{align}
    \func{\estm{B}_{\ell_1 \ell_2 L}}(k_1, k_2) = I_3^{-1} N_{\ell_1 \ell_2 L} H_{\ell_1 \ell_2 L} \sum_{m_1 m_2 M} & \wigthreej{\ell_1}{\ell_2}{L}{m_1}{m_2}{M} \notag \\
    & \times \int \dd[3]{\vx} \func{F_{\ell_1}^{m_1}}(\vx; k_1) \func{F_{\ell_2}^{m_2}}(\vx; k_2) \func{G_L^M}(\vx) \comma
    \label{eq:FFT-estimator-bispectrum} \\
    \func{\estm{\zeta}_{\ell_1 \ell_2 L}}(r_1, r_2) = I_3^{-1} N_{\ell_1 \ell_2 L} H_{\ell_1 \ell_2 L} \sum_{m_1 m_2 M} & \wigthreej{\ell_1}{\ell_2}{L}{m_1}{m_2}{M} \notag \\
    & \times \int \dd[3]{\vx} \func{F_{\ell_1}^{m_1}}(\vx; r_1) \func{F_{\ell_2}^{m_2}}(\vx; r_2) \func{G_L^M}(\vx) \comma
    \label{eq:FFT-estimator-3PCF}
\end{align}
\label{eq:FFT-estimators}%
\end{subequations}
with the normal{\is}ation constant
\begin{equation}
    I_3 \defeq \int \dd[3]{\vx} \func{w}(\vx)^3 \func{\nbar}(\vx)^3 \fstop
    \label{eq:normalisation-constant}
\end{equation}
Here the one-point functions
\begin{subequations}
\begin{align}
    \func{F_\ell^m}(\vx; k) &\defeq \int \frac{\dd[2]{\vuk}}{4\uppi} \e^{\im\vk\vdot\vx} \shyc{\ell}{m}\big(\vuk) \delta{n}(\vk) \comma \\
    \func{F_\ell^m}(\vx; r) &\defeq \im^\ell \int \frac{\dd[3]{\vk}}{(2\uppi)^3} \e^{\im\vk\vdot\vx} \func{j_\ell}(k r) \shyc{\ell}{m}\big(\vuk) \delta{n}(\vk) \comma \\
    \func{G_\ell^m}(\vx) &\defeq \int \frac{\dd[3]{\vk}}{(2\uppi)^3} \e^{\im\vk\vdot\vx} \func{\dn_\ell^m}(\vk)
\end{align}
\label{eq:one-point-fields}%
\end{subequations}
are obtained from \(\func{\dn_\ell^m}(\vk)\) either through binning in spherical shells in wavenumbers for \(\func{F_\ell^m}(\vx; k)\), or through inverse Fourier transforms of  \(\func{F_\ell^m}(\vx; r)\) and \(\func{G_\ell^m}(\vx)\).
The overall integration in \cref{eq:FFT-estimators} is computed as a sum of the product of these three one-point functions over all grid cells in configuration space.

Finally, to account for the discrete sampling of continuous number density fields by particles in a finite-sized catalogue, the shot noise components should be subtracted from the multipoles.
We shall use the expressions derived in ref.~\cite{Sugiyama:2019} (see eqs.~44--46 and 51 therein); however, we account for the mesh assignment window and aliasing effects in the shot components slightly differently, using the original exact prescription from ref.~\cite{Jing:2005} (see eqs.~19--21 and the discussion therein).

\subsection{Estimator of the three-point window function}

A key advantage of the TripoSH decomposition is that the window function for three-point clustering statistics,
\begin{multline}
    \func{Q}(\vr_1, \vr_2) = I_3^{-1} \int \dd[3]{\vx_1} \int \dd[3]{\vx_2} \int \dd[3]{\vx_3} \func{\dirac}(\vr_1 + \vx_1 - \vx_3) \func{\dirac}(\vr_2 + \vx_2 - \vx_3) \\
    \times \func{w}(\vx_1) \func{w}(\vx_2) \func{w}(\vx_3) \func{\nbar}(\vx_1) \func{\nbar}(\vx_2) \func{\nbar}(\vx_3) \comma
\end{multline}%
can be decomposed in exactly the same basis as for the bispectrum or 3PCF, with multipoles given by
\begin{align}
    \func{Q_{\ell_1 \ell_2 L}}(r_1, r_2) &= I_3^{-1} N_{\ell_1 \ell_2 L} H_{\ell_1 \ell_2 L} \sum_{m_1 m_2 M} \wigthreej{\ell_1}{\ell_2}{L}{m_1}{m_2}{M} \int \frac{\dd[2]{\vur_1}}{4\uppi} \func{\shyc{\ell_1}{m_1}}\big(\vur_1) \frac{\dd[2]{\vur_2}}{4\uppi} \func{\shyc{\ell_2}{m_2}}\big(\vur_2) \notag \\
    & \mquad[4] \times \int \dd[3]{\vx_1} \int \dd[3]{\vx_2} \int \dd[3]{\vx_3} \func{\dirac}(\vr_1 + \vx_1 - \vx_3) \func{\dirac}(\vr_2 + \vx_2 - \vx_3) \notag \\ 
    & \mquad[8] \times \func{\shyc{L}{M}}(\vux_3) \func{w}(\vx_1) \func{w}(\vx_2) \func{w}(\vx_3) \func{\nbar}(\vx_1) \func{\nbar}(\vx_2) \func{\nbar}(\vx_3) \comma
    \label{eq:3PCF-multipole-window}
\end{align}%
where \(\dirac\) denotes the Dirac delta function.
From these multipoles, a straightforward window convolution series can be derived as shown in the next section.
In configuration space, the window function multipoles~\(\func{Q_{\ell_1 \ell_2 L}}(r_1, r_2)\) can be estimated from dense random catalogues using \cref{eq:FFT-estimator-3PCF} derived for the 3PCF by replacing \(\dn\) with \(w \nbar\) in \cref{eq:one-point-fields}.
Note that under the simultaneous exchange \(\ell_1 \leftrightarrow \ell_2\) and \(r_1 \leftrightarrow r_2\), the value of TripoSH multipoles remains the same.
This exchange symmetry means that for multipoles with degrees \(\ell_1 = \ell_2\), only the upper triangular part (\(r_1 \leq r_2\)) of the two-dimensional quantity \(\func{Q_{\ell_1 \ell_2 L}}(r_1, r_2)\) needs to be computed.

By contrast, in other approaches such as ref.~\cite{Pardede:2022}, the window function multipoles have a complex form with a larger number of degrees of freedom (namely additional spherical degree~\(\ell\) and order~\(m\) indices) than the three-point clustering statistics themselves and thus need computationally efficient estimators to be separately developed.
In this aspect, window convolution in the TripoSH formalism resembles the standard window convolution procedure for two-point clustering statistics much more closely~\cite{Wilson:2016,Beutler:2016}.

The estimation of the three-point window function multipoles and the shot noise components are both implemented by our publicly available code \codename{Triumvirate}~\cite{Wang:2023a} introduced in \cref{sec:introduction}.

\section{Window convolution recipe}
\label{sec:convolution}

In this section, we will briefly review the window convolution formula given in ref.~\cite{Sugiyama:2019} and then discuss the practical implementation of the window convolution procedure, including a linear algebra formulation that provides a window matrix for premultiplication with the bispectrum model vector.

\subsection{Window convolution series}

Since the true background number density is unknown and has to be inferred from observations, this leads to an integral constraint placed on the over-density field and its correlators~\cite{Peacock:1991};
consequently, this needs to be corrected for in any observable 3PCF, \(\zeta \mapsto \zeta - \zetabar\).
In this work, we adopt the treatment from ref.~\cite{Sugiyama:2019} by assuming the correction~\(\zetabar\) is a constant determined from the isotropic condition
\begin{equation}
    \int \dd{r_1} r_1^2 \dd{r_2} r_2^2 \zeta_{000}(r_1, r_2) = 0 \fstop
\end{equation}
More generally, this correction should be scale-dependent as the constraint placed on the over-density field couples with the field itself, as demonstrated for two-point clustering statistics in ref.~\cite{deMattia:2019}.
In addition, here we also ignore any radial integral constraint that arises when the radial selection function also has to be inferred from the survey observations directly.
We leave a more formal and consistent study of the integral constraint corrections in three-point clustering statistics to future work.

To proceed, we consider the following 3PCF multipole estimator that is mathematically equivalent to \cref{eq:FFT-estimator-3PCF}:
\begin{align}
    \func{\estm{\zeta}_{\ell_1 \ell_2 L}}(r_1, r_2) &= I_3^{-1} N_{\ell_1 \ell_2 L} H_{\ell_1 \ell_2 L} \sum_{m_1 m_2 M} \wigthreej{\ell_1}{\ell_2}{L}{m_1}{m_2}{M} \int \frac{\dd[2]{\vur_1}}{4\uppi} \func{\shyc{\ell_1}{m_1}}\big(\vur_1) \frac{\dd[2]{\vur_2}}{4\uppi} \func{\shyc{\ell_2}{m_2}}\big(\vur_2) \notag \\
    & \mquad[4] \times \int \dd[3]{\vx_1} \int \dd[3]{\vx_2} \int \dd[3]{\vx_3} \func{\dirac}(\vr_1 + \vx_1 - \vx_3) \func{\dirac}(\vr_2 + \vx_2 - \vx_3) \notag \\ 
    & \mquad[8] \times \func{\shyc{L}{M}}(\vux_3) \func{\dn}(\vx_1) \func{\dn}(\vx_2) \func{\dn}(\vx_3) \fstop
    \label{eq:3PCF-multipole-estimator}
\end{align}
By taking the ensemble average of this estimator and substituting \cref{eq:3PCF-local-pp} with the integral constraint correction~\(\zeta \mapsto \zeta - \zetabar\) and the TripoSH multipole expansion, one obtains the window-convolved 3PCF model~\cite{Sugiyama:2019}
\begin{multline}
    \expval{\func{{\estm{\zeta}_{\ell_1 \ell_2 L}}}(r_1, r_2)} = N_{\ell_1 \ell_2 L} \sum_{\ell'_1 \ell'_2 L'} \sum_{\ell''_1 \ell''_2 L''} \wigninej{\ell''_1}{\ell''_2}{L''}{\ell'_1}{\ell'_2}{L'}{\ell_1}{\ell_2}{L} \frac{H_{\ell_1 \ell'_1 \ell''_1} H_{\ell_2 \ell'_2 \ell''_2} H_{L L' L''} H_{\ell_1 \ell_2 L}}{H_{\ell'_1 \ell'_2 L'} H_{\ell''_1 \ell''_2 L''}} \\
    \times \func{{Q_{\ell''_1 \ell''_2 L''}}}(r_1, r_2) \func{{\zeta_{\ell'_1 \ell'_2 L'}}}(r_1, r_2) - \func{{Q_{\ell_1 \ell_2 L}}}(r_1, r_2) {\zetabar} \fstop
    \label{eq:window-convolution-series}
\end{multline}
Here the window function multipoles~\(\func{Q_{\ell_1 \ell_2 L}}(r_1, r_2)\) are given by \cref{eq:3PCF-multipole-estimator} with the replacement of \(\func{\dn}(\vx)\) by \(\func{w}(\vx) \func{\nbar}(\vx)\), and the integral constraint correction is given by
\begin{equation}
    \zetabar = \expval{Q_{000}, 1}^{-1} \sum_{\ell_1 \ell_2 L} \frac{\expval{Q_{\ell_1 \ell_2 L}, \zeta_{\ell_1 \ell_2 L}}}{N_{\ell_1 \ell_2 L} H_{\ell_1 \ell_2 L}^{2}} \comma
    \label{eq:integral-constraint-series}
\end{equation}
where we have defined
\begin{equation}
    \expval{A, B} \defeq \int \dd{r_1} r_1^2 \dd{r_2} r_2^2 \func{A}(r_1, r_2) \func{B}(r_1, r_2) \fstop
\end{equation}

In general, \cref{eq:window-convolution-series,eq:integral-constraint-series} are infinite series which have to be truncated in practical evaluations.
As the multipole degree \(L\), \(L'\) or \(L''\) associated with line-of-sight anisotropies increases, the amplitude of the corresponding multipole generally decreases, but this relation does not necessarily hold for \(\ell_{1,2}\); as such, for each survey geometry, the convergence of \cref{eq:window-convolution-series} or \eqref{eq:integral-constraint-series} needs to be validated by empirically determining which terms in the series make a significant contribution.
This is what we will verify in \cref{sec:validation}.

\subsection{Double spherical Bessel transform}

The double spherical Bessel transform (eqs.~\ref{eq:double-spherical-Bessel-transform}) required to convert between bispectrum and 3PCF multipoles is numerically challenging because of the oscillatory nature of the spherical Bessel functions in the integrand. 
The same type of transform, however, also appears in the transform between power spectrum and two-point correlation function multipoles and is dealt with using the FFTLog algorithm~\cite{Talman:1978,Hamilton:2000,Wilson:2016,Karamanis:2021}.
In this work, we take the same approach as in ref.~\cite{Sugiyama:2019} by treating the double spherical Bessel transform as two sequential one-dimensional Hankel transforms performed using FFTLog.
In the future, one may wish to consider alternative numerical algorithms such as 2D-FFTLog~\cite{Fang:2020}, Levin integration~\cite{Levin:1996,Levin:1997,Leonard:2023} and a complex integration method based on the Picard--Lefschetz theory~\cite{Feldbrugge:2023}.

If one is interested in a 3PCF analysis, then only the backward double spherical Bessel transform of a bispectrum model into configuration space is needed. However, the bispectrum multipoles must be specified in the full two-dimensional form, \(\func{B_{\ell_1 \ell_2 L}}(k_1, k_2)\), even when the 3PCF multipoles are restricted to the diagonal form, \(\func{\zeta_{\ell_1 \ell_2 L}}(r, r)\).
On the other hand, only the diagonal part of the window function multipoles, \(\func{Q_{\ell_1 \ell_2 L}}(r, r)\), is needed as the convolution is locally multiplicative in \cref{eq:window-convolution-series}.
By contrast, a bispectrum analysis even when restricted to the diagonal part~\(\func{B_{\ell_1 \ell_2 L}}(k, k)\) requires the full two-dimensional window function multipoles~\(\func{Q_{\ell_1 \ell_2 L}}(r_1, r_2)\), since additional forward double spherical Bessel transforms are required after window convolution is performed in configuration space.

\subsection{Linear algebra formulation}
\label{sec-sub:convolution-matrix-formulation}

Following the discussion in the previous subsections, the full window convolution recipe for a bispectrum analysis can be summarised by the steps below.
\begin{enumerate}
    \item Measure the full two-dimensional window function multipoles~\(\func{Q_{\ell_1 \ell_2 L}}(r_1, r_2)\) from a random catalogue that samples the survey geometry (applying eq.~\ref{eq:FFT-estimator-3PCF}).
    \item Transform a given bispectrum model for all multipoles~\(\func{B_{\ell_1 \ell_2 L}}(k_1, k_2)\) required by the (truncated) window convolution series into 3PCF multipoles~\(\func{\zeta_{\ell_1 \ell_2 L}}(r_1, r_2)\) (applying eq.~\ref{eq:double-spherical-Bessel-transform-backward}).
    \item Evaluate the window convolution series to obtain the window-convolved 3PCF multipoles~\(\func{\wind{\zeta}_{\ell_1 \ell_2 L}}(r_1, r_2)\) (applying eq.~\ref{eq:window-convolution-series}).
    \item Transform the window-convolved 3PCF multipoles back into desired window-convolved bispectrum multipoles~\(\func{\wind{B}_{\ell_1 \ell_2 L}}(k_1, k_2)\) (applying eq.~\ref{eq:double-spherical-Bessel-transform-forward}).
\end{enumerate}
The truncation of the window convolution series~\cref{eq:double-spherical-Bessel-transform-backward} to achieve convergence depends on the specific survey geometry and needs to be verified for each analysis as we will demonstrate in \cref{sec:validation}.

It transpires that each step above is a linear algebra operation: the double spherical Bessel transform is a linear integral transform, and the window convolution series including the integral constraint is a linear combination of unwindowed model multipoles.
Therefore, the full window convolution procedure can be recast as a single linear algebra operation; schematically, \(\vwB = \mW \vB\), where \(\vB\) and \(\vwB\) are the unwindowed and windowed bispectrum model vectors and \(\mW\) is the window matrix that captures the full window convolution procedure.

Although the elements of the window matrix~\(\mW\) can be explicitly and individually computed, one may rely on the fact that for a linear transformation, the columns of the transformation matrix correspond to the transformed basis vectors.
Therefore, one could construct the window matrix~\(\mW\) as follows, provided the wavenumbers~\(\{k_i\}_{i=1}^{N_k}\) at which the bispectrum model is evaluated are predefined. First, 
\begin{enumerate}
    \item fix the ordering of multipoles, e.g. sort the multipole degrees~\(\{(\ell_1, \ell_2, L)\}\) by \(\ell_{1,2}\) and then by \(L\) overall;
    \item vector{\is}e the two-dimensional bispectrum multipoles in row-major order, i.e. \(\func{B_{\ell_1 \ell_2 L}}(k_i, k_j)\) is the \([N_k (i-1) + j]\)-th element of the (unwindowed) multipole vector~\(\vB_{\ell_1 \ell_2 L}\);
    \item concatenate the required unwindowed multipole vectors (say \(N_\mathrm{pole}\) in total) to form the full unwindowed model vector~\(\trs{\vB} = \big(\trs{\vB}_{\ell_1 \ell_2 L}, \dots\big)\) in the fixed order.
\end{enumerate}
Next,
\begin{enumerate}[resume]
    \item feed the basis of unit vectors~\(\vb{u}_I = \trs{(0, \dots, 1, \dots, 0)}\), where the index~\(I = 0, \dots, N_\mathrm{pole} N_k^2\), as \(\vB\) into the full window convolution procedure; in other words, each time set \(\func{B_{\ell'_1 \ell'_2 L'}}(k_{i'}, k_{j'}) = 1\) for one particular multipole~\((\ell'_1, \ell'_2, L')\) at specific wave\-numbers \((k_{i'}, k_{j'})\), and all other multipole values to zero;
    \item for the \(I\)-th unit vector, concatenate the resulting windowed multipole vectors into \(\trs{\vwu} = \big(\trs{\vwu}_{\ell_1 \ell_2 L}, \dots\big)\) in the fixed order, and this is the \(I\)-th column of the window matrix~\(\mW\).
\end{enumerate}
Now for any unwindowed bispectrum model evaluated at the same wavenumbers and vector{\is}ed as above, one could premultiply the model vector~\(\vB\) by the same window matrix~\(\mW\) to obtain a windowed bispectrum model vector~\(\vwB\) that can be compared with the bispectrum measurements vector{\is}ed in the same way, e.g. in a likelihood analysis.

\section{Validation tests}
\label{sec:validation}

Having laid down the full procedure for the window convolution of the TripoSH bispectrum multipoles, we now validate the pipeline (including the linear algebra formulation) with the Dark Energy Spectroscopic Instrument~(DESI) Data Release~1~(DR1) survey specification~\citep{Levi:2013,DESI2016.Science,DESI2016.Instr,DESI2022.KP1.Instr,DESI2024.KP1.SV,DESI2024.KP1.EDR,DESI2024.I.DR1,DESI2024.II.KP3,DESIL2024.III.KP4,DESI2024.IV.KP6,DESI2024.VII.KP7,Silber:2023,Miller:2023,Guy:2023,Schlafly:2023,Zhou:2023}.
We will focus on the luminous red galaxy~(LRG) sample in the South Galactic Cap~(SGC) in the redshift bin \(\num{0.4} \leq z \leq \num{0.6}\) in the DR1 data~\cite{DESI2024.I.DR1,DESI2024.II.KP3}, and consider the diagonal windowed bispectrum monopole~\(\func{\wind{B}_{000}}(k, k)\) and quadrupole~\(\func{\wind{B}_{202}}(k, k)\) which have the highest signal-to-noise ratio~\cite{Sugiyama:2019}.
This LRG sample in the lowest redshift bin in the SGC region is chosen since it has a relatively small comoving volume (and a similar effective volume) of \SI{0.5}{\cubic\per\h\cubic\giga\parsec}~\cite{DESI2024.II.KP3,DESI2024.V.KP5}; therefore, the survey window is expected to have a significant impact, which could make the window convolution procedure more challenging to deal with.

To perform the validation tests, we make use of the DESI Second-Generation mock catalogues based on \num{25} real{\is}ations of the \codename{AbacusSummit} simulation suite in the baseline Planck~2018 cosmology~\cite{Maksimova:2021,Garrison:2021,Planck18:2020b},
which have been calibrated against the two-point clustering characteristics of the DESI One-Percent Survey~\cite{Yuan:2024}.
These mock catalogues include fixed-redshift simulation snapshots in \((\SI{2}{\per\h\giga\parsec})^3\) cubic boxes (at \(z = \num{0.5}\) for the sample considered here), as well as cut-sky catalogues (including random catalogues) replicated from the snapshots to match the DR1 survey footprint including the veto mask and the number density redshift distribution, \(\func{\nbar}(z)\).

We will use the mean of the \num{25} cubic-box measurements as a proxy bispectrum model without window effects since they share exactly the same underlying clustering signal as the cut-sky mock catalogues on scales below the box size.%
\footnote{The cut-sky mock catalogues are not light cones and thus have no redshift evolution.}
This avoids dependence on a specific theoretical model, which may have a limited scale range of predictive accuracy, and thus helps isolate and diagnose any discrepancies arising from the window convolution treatment itself.
We measure the three-point window function from the cut-sky random catalogues and compare any window-convolved bispectrum multipoles with the mean of the \num{25} cut-sky measurements taken as the reference.
For the full pipeline validation, the cut-sky mock catalogues used have been generated from the alternate merged target ledgers (AMTL)~\cite{Lasker:2024}, which realistically simulate the fibre assignment procedure of actual DESI observations.
However, we focus solely on the survey window effect here and do not discuss any potential impact of fibre assignment.
The shot noise component in the cut-sky clustering measurements is also affected by the survey window, and may be modelled with additional nuisance parameters at the residual level (after accounting for the standard Poissonian shot noise) and then window-convolved in cosmological analyses~\cite[e.g.][]{Pardede:2022}.
Instead, here we choose to consistently subtract the estimated shot noise in all measurements including the cubic-box ones taken as the proxy model, since the number density is very different between the cubic-box and cut-sky mock catalogues owing to downsampling.

\subsection{Window function measurements}
\label{sec-sub:validation-window}

On scales larger than the catalogue extent, the window function multipoles~\(\func{Q_{\ell_1 \ell_2 L}}(r_1, r_2)\) should be identically zero; however, the relatively limited number of wavevector modes on the largest scales means that sample variance may manifest itself as a non-vanishing multipole value.
Therefore, the random catalogue should be assigned to a mesh grid that is appreciably larger than its extent for FFT sampling. 

At the other end, the FFT measurements suffer from aliasing effects local{\is}ed to small scales in Fourier space (which may spread over a wider range of scales in configuration space).
This can be mitigated by using a higher-resolution mesh grid, which means a substantial number of grid cells when the physical size of the mesh grid is large.
As such, the requirement to measure three-point window function multipoles over a large dynamic range imposes a significant computational demand.

For the two-point window function of a single scale variable, one may circumvent this issue by repeating the measurement over two mesh grids separately: one larger and coarser for long scales, and one smaller and finer for short scales.
The measurements should agree well on intermediate scales and can thus be concatenated.
Unfortunately, the fact that the three-point window function multipole~\({Q_{\ell_1 \ell_2 L}}\) is a function of two scale variables means that this would not work for its off-diagonal elements.
Instead, in \cref{sec-subsub:validation-window-convergence-mesh} we measure the window function using the largest mesh grid computationally feasible in terms of physical size and cell numbers, and verify that the measurements have already converged with smaller mesh grid dimensions and cell numbers.

A separate issue with small-scale measurements of the window function arises when the random catalogue is not sufficiently dense so that the short-scale modes below the mean inter-particle separation are poorly sampled.
In \cref{sec-subsub:validation-window-convergence-density}, we also perform convergence tests by varying the number density of the random catalogue.

After finding the measurement configuration that achieves percent-level convergence, we measure the set of window function multipoles used for testing the window convolution series in \cref{sec-sub:validation-series}.
Some of these multipoles are shown in \cref{fig:window-function-multipoles} at the end of \cref{sec-subsub:validation-window-convergence-density}.

\subsubsection{Convergence with mesh sampling}
\label{sec-subsub:validation-window-convergence-mesh}

\begin{figure}
    \centering
    \includegraphics[width=0.75\linewidth]{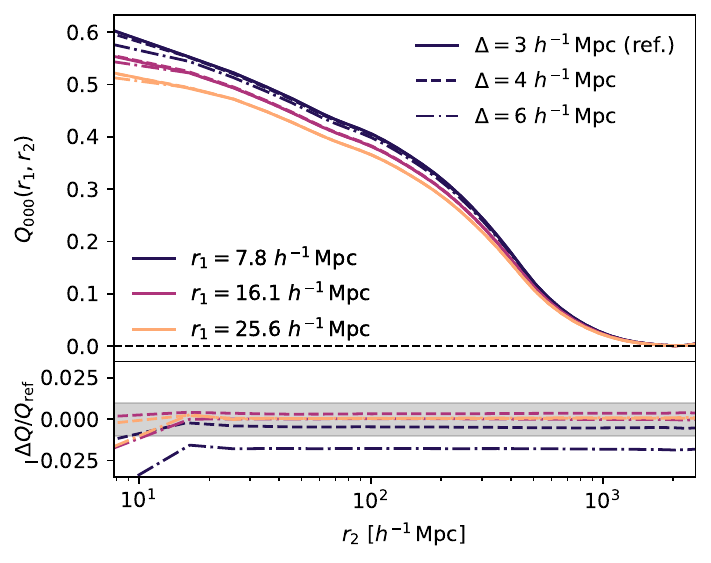}
    \caption{%
        Three-point window function monopole~\(\func{Q_{000}}(r_1, r_2)\) for the DESI DR1 LRG SGC sample in the redshift bin \(\num{0.4} \leq z \leq \num{0.6}\), measured with different grid cell sizes to check convergence with respect to aliasing effects.
        The \textit{top panel} shows the measurements at multiple small fixed values~\(r_1 = \num{7.8}\), \num{16.1} and \SI{25.6}{\lunit} with a varying mesh grid cell size~\(\Delta = \num{3}\), \num{4} and \SI{6}{\lunit}.
        The \textit{bottom panel} shows the relative difference of the measurements with respect to the reference case of \(\Delta = \num{3}~\si{\lunit}\), where the grey shaded region delimits \SI{\pm 1}{\percent} deviations.
    }
    \label{fig:window-convergence-fftsamp-cellsize}
\end{figure}

We first measure the window function monopole~\(\func{Q_{000}}(r_1, r_2)\) from a mesh grid of size \((L_x, L_y, L_z) = (\num{3072}, \num{6144}, \num{3072})~\si{\lunit}\), which is at least twice the extent of the random catalogue in each dimension.
The separation~\(r_{1,2}\) is chosen to be binned linearly for \(r \in (0, \num{50}]~\si{\lunit}\) and logarithmically for \(r \in (\num{50}, \num{4000}]~\si{\lunit}\).
In \cref{fig:window-convergence-fftsamp-cellsize}, we compare the measurements at multiple small fixed values of \(r_1\) for a varying grid cell size~\(\Delta\).%
\footnote{%
For fixed mesh grid dimensions, varying the cell size is equivalent to varying the cell number.
Values of \(r_1\) are the effective separation obtained after averaging in each bin.
}
As the grid cell size approaches \(\Delta = \num{3}~\si{\lunit}\), the measurement of \(\func{Q_{000}}(r_1, r_2)\) has converged to within about \SI{1}{\percent}, and hereafter this will be used as the maximum grid cell size when measuring the three-point window function.

Next, with the grid cell size fixed at \(\Delta = \num{3}~\si{\lunit}\), we measure the diagonal window function monopole~\(\func{Q_{000}}(r, r)\) in the same separation bins for multiple mesh grid sizes \((L_x, L_y, L_z) = (L, 2L, L)\) where \(L\) is varied.
\begin{figure}
    \centering
    \includegraphics[width=0.75\linewidth]{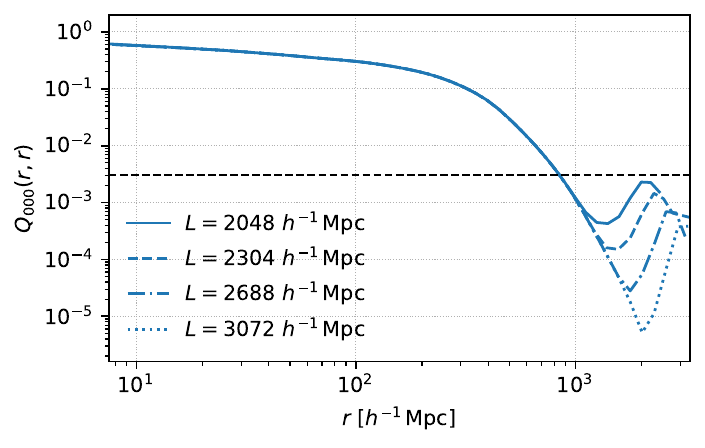}
    \caption{%
        Diagonal three-point window function monopole~\(\func{Q_{000}}(r, r)\) for the DESI DR1 LRG SGC sample in the redshift bin \(\num{0.4} \leq z \leq \num{0.6}\), measured with different mesh grid sizes to check the effect of sample variance of large-scale modes.
        Each line shows the measurement with a different mesh grid size \((L_x, L_y, L_z) = (L, 2L, L)\), where \(L = \num{2048}\), \num{2304}, \num{2688} and \SI{3072}{\lunit} respectively.
        The horizontal dashed line marks \SI{0.5}{\percent} of the maximum value, i.e. \(\num{0.005} \max_{r} \func{Q_{000}}(r, r)\).
    }
    \label{fig:window-convergence-fftsamp-meshdim}
\end{figure}
In \cref{fig:window-convergence-fftsamp-meshdim}, we see that as \(r\) increases, the window function multipole becomes vanishingly small, albeit with a nonzero fluctuating amplitude due to the sample variance of large-scale modes; however, with a greater mesh grid size, the sample variance is reduced and so is the window function residual.
For the LRG SGC sample in the redshift bin \(\num{0.4} \leq z \leq \num{0.6}\) considered in this section, the mesh grid size with \(L = \SI{2688}{\lunit}\) is adopted, as the residual amplitude of \(\func{Q_{000}}(r, r)\) for large \(r\) is suppressed to below about \SI{0.1}{\percent} of its maximum.

\subsubsection{Convergence with number density}
\label{sec-subsub:validation-window-convergence-density}

We also test the convergence of the window function measurements with respect to the number density of the random catalogue.
Typically, the number density of the random catalogue is expressed as an inverse ratio of weighted number counts compared with that of the data catalogue, \(\alpha \defeq \qty(\sum_\mathrm{data} w_\sys) / \sum_\mathrm{random} w_\sys\), where \(w_\sys\) is the systematic weight (unity if not applied).
\begin{figure}
    \centering
    \includegraphics[width=0.75\linewidth]{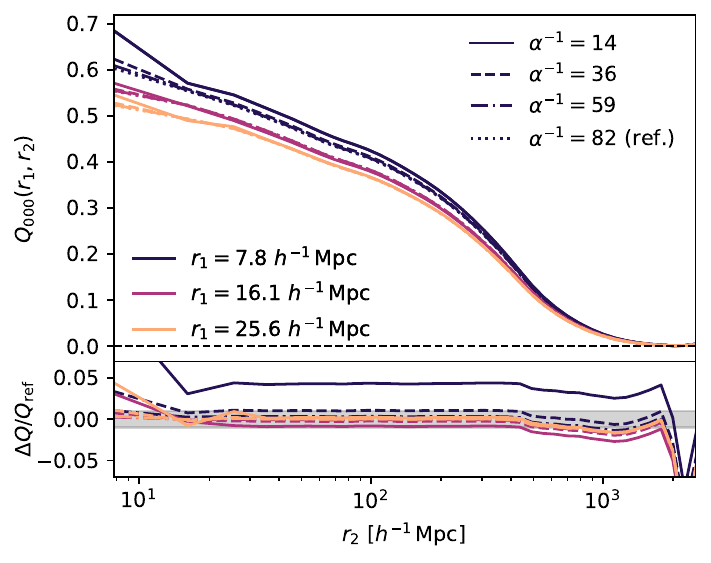}
    \caption{%
        Three-point window function monopole~\(\func{Q_{000}}(r_1, r_2)\) for the DESI DR1 LRG SGC sample in the redshift bin \(\num{0.4} \leq z \leq \num{0.6}\).
        The top panel\textit{} shows the measurements at multiple small fixed values~\(r_1 = \num{7.8}\), \num{16.1} and \SI{25.6}{\lunit} with a varying random catalogue number density ratio~\(\alpha^{-1} = \num{14}\), \num{36}, \num{59} and \num{82}.
        The \textit{bottom panel} shows the relative difference of the measurements with respect to the reference case of \(\alpha^{-1} = \num{82}\), where the grey shaded region delimits \SI{\pm 1}{\percent} deviations.
    }
    \label{fig:window-convergence-numsamp}
\end{figure}
In \cref{fig:window-convergence-numsamp}, we compare the measurements at multiple small fixed values of \(r_1\) for a varying ratio~\(\alpha^{-1}\).
As the random catalogue becomes denser with an increasing value of \(\alpha^{-1}\), the measurement of \(\func{Q_{000}}(r_1, r_2)\) has converged to within about \SI{1}{\percent}.
Given that the increased number density of the random catalogue has a negligible impact on computation time, hereafter we will always use the densest random catalogue available for all measurements.

We briefly note here that the number density of the random catalogue appears to have an effect on the largest scales, but there is no discernible trend with \(\alpha^{-1}\).
We attribute this also to the sample variance of the large-scale modes, as the random catalogues at different number densities are fundamentally different real{\is}ations.
This effect is visually more pronounced in the bottom panel of \cref{fig:window-convergence-numsamp} showing the relative deviation, because the amplitude of \(\func{Q_{000}}(r_1, r_2)\) is itself small and close to zero for large \(r_2\); the absolute deviation remains small.

\begin{figure}
    \centering
    \includegraphics[width=0.75\linewidth]{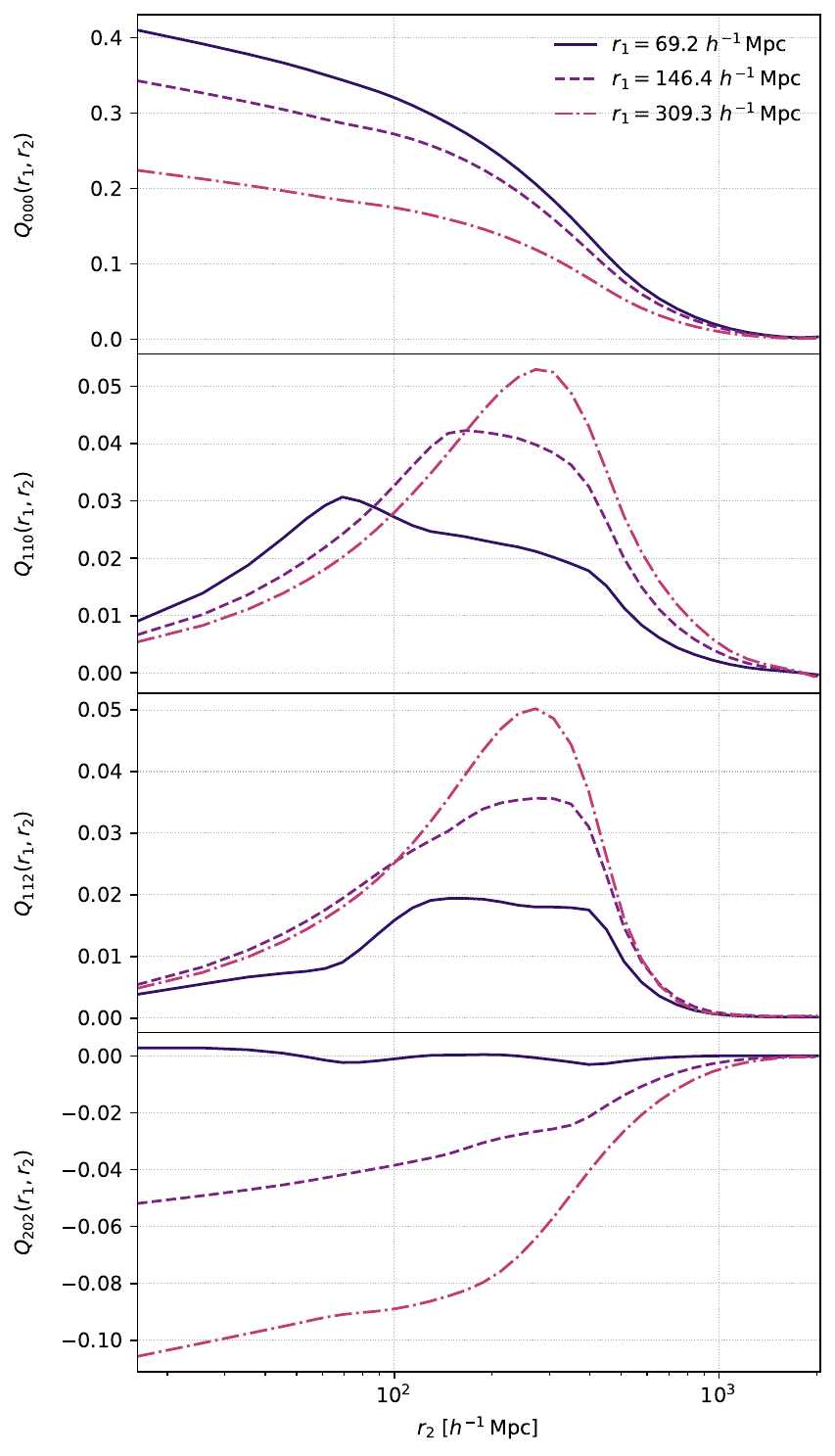}
    \caption{%
        Some of the three-point window function multipoles~\(\func{Q_{\ell_1 \ell_2 L}}(r_1, r_2)\) for the DESI DR1 LRG SGC sample in the redshift bin \(\num{0.4} \leq z \leq \num{0.6}\), with \(r_1\) fixed at multiple different values.
    }
    \label{fig:window-function-multipoles}
\end{figure}

Having tested the convergence of window function measurements, in \cref{fig:window-function-multipoles} we show some of the three-point window function multipoles to be used in the convolution series in \cref{sec-sub:validation-series}.
These measurements have been made from the densest random catalogue available (\(\alpha^{-1} = \num{82}\)) with a mesh grid of minimum dimension~\(L = \SI{2688}{\lunit}\) and cell size~\(\Delta = \SI{2.6}{\lunit}\).

\subsection{Window convolution series}
\label{sec-sub:validation-series}

In this subsection, we examine the contribution of the various terms in the window convolution series~\eqref{eq:window-convolution-series} for the bispectrum monopole~\(\wind{B}_{000}\) and quadrupole~\(\wind{B}_{202}\).
We consider the following truncated series with a large number of 3PCF multipoles as the reference formula,
\begin{subequations}
\begin{align}
\wind{\zeta}_{000} & =
Q_{000} \zeta_{000} \notag \allowdisplaybreaks \\
& \qquad + \frac{1}{3} Q_{110} \zeta_{110} + \frac{1}{5} Q_{220} \zeta_{220} \notag  \allowdisplaybreaks \\
& \qquad + \frac{1}{5} \qty(Q_{022} \zeta_{022} + Q_{202} \zeta_{202}) + \frac{1}{6} Q_{112} \zeta_{112} + \frac{1}{9} \qty(Q_{132} \zeta_{132} + Q_{312} \zeta_{312}) \notag \allowdisplaybreaks \\
& \qquad - Q_{000} \zetabar \comma \allowdisplaybreaks \\
\wind{\zeta}_{202} & =
Q_{000} \zeta_{202} + Q_{202} \zeta_{000} \notag \allowdisplaybreaks \\
& \qquad + \frac{1}{3} \qty(Q_{112} \zeta_{110} + Q_{312} \zeta_{110} + Q_{110} \zeta_{112} + Q_{110} \zeta_{312}) + \frac{1}{5} \qty(Q_{022} \zeta_{220} + Q_{220} \zeta_{022}) \notag \allowdisplaybreaks \\
& \qquad + \frac{2}{7} Q_{202} \zeta_{202} + \frac{1}{6} \qty(Q_{112} \zeta_{112} + Q_{132} \zeta_{132}) + \frac{8}{63} Q_{312} \zeta_{312} + \frac{1}{21} \qty(Q_{312} \zeta_{112} + Q_{112} \zeta_{312}) \notag \allowdisplaybreaks \\
& \qquad - Q_{202} \zetabar \fstop
\end{align}
\label{eq:window-convolution-formula-ref}%
\end{subequations}
Here the set of input multipoles are \((\ell_1, \ell_2, L) \in \{(0, 0, 0),\allowbreak (1, 1, 0),\allowbreak (2, 2, 0),\allowbreak (2, 0, 2),\allowbreak (1, 1, 2),\allowbreak (1, 3, 2)\}\) plus their counterparts under the exchange \(\ell_1 \leftrightarrow \ell_2\), which include monopoles and quadrupoles with high signal-to-noise ratios.
The integral constraint correction given by the series~\eqref{eq:integral-constraint-series} is calculated from the same set of multipoles.

To obtain the set of unwindowed 3PCF multipoles~\(\func{\zeta_{\ell_1 \ell_2 L}}(r_1, r_2)\), we double spherical Bessel transform the two-dimensional proxy model~\(\func{B_{\ell_1 \ell_2 L}}(k_1, k_2)\) in \num{50} logarithmic bins of wavenumber \(k \in [0.001, 0.520]~\si{\wunit}\) in each dimension.
The window function and unwindowed 3PCF multipoles are resampled over the same logarithmically spaced points wherever necessary for the multiplication in \cref{eq:window-convolution-series} and for the FFTLog algorithm.
Finally, we double spherical Bessel transform the window-convolved 3PCF multipoles~\(\func{\wind{\zeta}_{\ell_1 \ell_2 L}}(r_1, r_2)\) back into Fourier space to obtain the window-convolved bispectrum multipoles~\(\func{\wind{B}_{\ell_1 \ell_2 L}}(k_1, k_2)\) as described in \cref{sec:convolution}.

\begin{figure}
    \centering
    {\small DESI DR1 LRG SGC \(\num{0.4} \leq z \leq \num{0.6}\)\smallskip}
    \includegraphics[width=\linewidth]{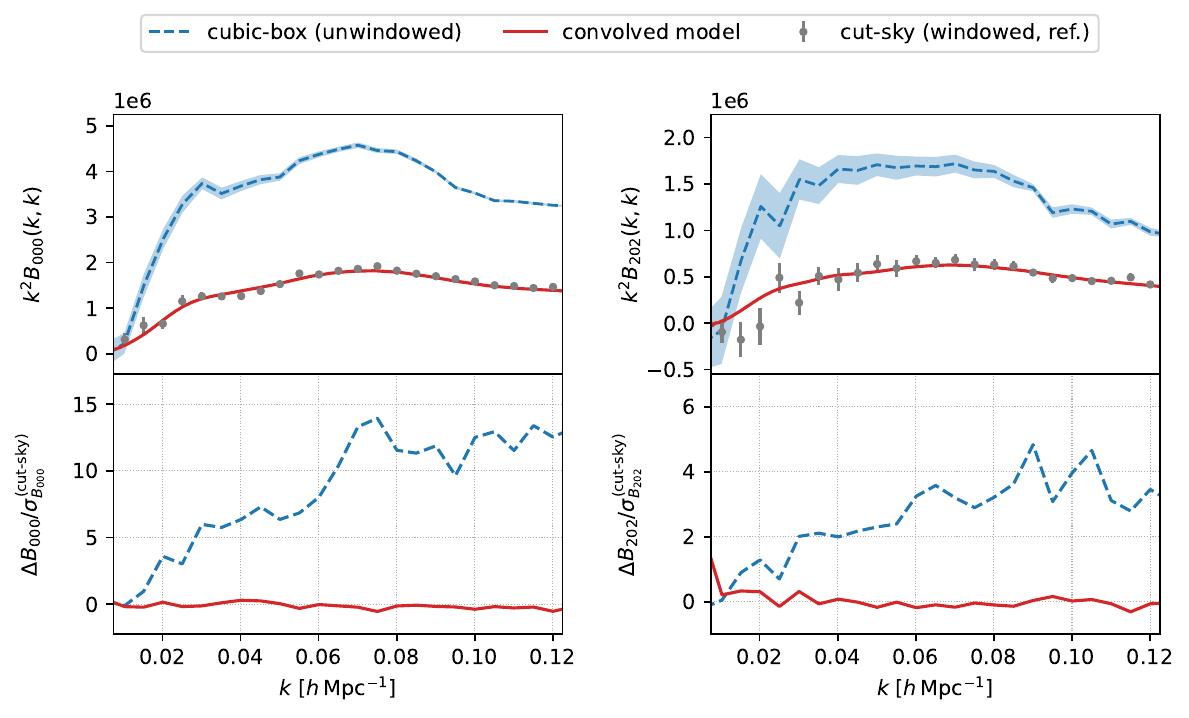}
    \caption{%
        The \textit{top panels} compare the diagonal bispectrum monopole~\(\func{B_{000}}(k, k)\) (\textit{left column}) and quadrupole~\(\func{B_{202}}(k, k)\) (\textit{right column}) of unwindowed cubic-box measurements, windowed cut-sky measurements and the window-convolved proxy model.
        The \textit{bottom panels} show the deviation~\(\Delta{B}_{\ell_1 \ell_2 L}\) of the window-convolved proxy model from the cut-sky measurements as a multiple of the standard deviation~\(\sigma_{B_{\ell_1 \ell_2 L}}^\text{(cut-sky)}\) of the cut-sky measurements. 
    }
    \label{fig:window-convolution-full}
\end{figure}

In \cref{fig:window-convolution-full}, we compare the window-convolved bispectrum proxy model~\(\wind{B}_{\ell_1 \ell_2 L}\) with the windowed cut-sky measurements and the unwindowed cubic-box measurements in the wavenumber range~\((0, 0.12]~\si{\wunit}\). 
As mentioned earlier, we focus on the diagonal monopole and quadrupole which retain the most amount of signal relative to noise.
We find that the window function significantly changes the amplitude and shape of the bispectrum multipoles; without accounting for its effect, there can be a deviation of up to \SI{14}{\stdsig} in the monopole~\(\wind{B}_{000}\) and up to \SI{5}{\stdsig} in the quadrupole~\(\wind{B}_{202}\), where \(\sigma\) is the standard deviation of the windowed cut-sky measurements.
By contrast, the window-convolved bispectrum proxy model has no noticeable deviation from the cut-sky measurements.

As a useful diagnostic tool, we consider the possibility of a constant amplitude offset of the window-convolved bispectrum proxy model from the windowed cut-sky measurements, which is parametr{\is}ed as a relative difference, \(\wind{B}_{\ell_1 \ell_2 L} \mapsto (1 + \beta) \wind{B}_{\ell_1 \ell_2 L}\). 
To determine the amplitude offset parameter~\(\beta\) as well as the contribution of the various terms in the window convolution series~\eqref{eq:window-convolution-series}, we define the following loss function:
\begin{multline}
    \func{\chi^2_{\ell_1 \ell_2 L}}(\Lambda, \beta) = \sum_{i,j} \qty[(1 + \beta) \func{\wind{B}_{\ell_1 \ell_2 L}^{(\Lambda)}}(k_i, k_i) - \func{B_{\ell_1 \ell_2 L}^\text{(cut-sky)}}(k_i, k_i)] \\
    \times \qty(\mat{C}_{\ell_1 \ell_2 L}^{-1})_{ij} \qty[(1 + \beta) \func{\wind{B}_{\ell_1 \ell_2 L}^{(\Lambda)}}(k_j, k_j) - \func{B_{\ell_1 \ell_2 L}^\text{(cut-sky)}}(k_j, k_j)] \fstop
    \label{eq:loss-function}
\end{multline}
Here the diagonal window-convolved bispectrum proxy model~\(\func{\wind{B}_{\ell_1 \ell_2 L}^{(\Lambda)}}(k, k)\) depends on the window convolution formula labelled by \(\Lambda\), and \(\mat{C}_{\ell_1 \ell_2 L}\) is the covariance matrix of the diagonal windowed cut-sky bispectrum multipole~\(\func{B_{\ell_1 \ell_2 L}^\text{(cut-sky)}}(k, k)\) estimated from 25 measurements.
These quantities are evaluated at the effective wavenumber~\(k_i\) in \num{12} (\(\ll 25\)) bins in the range mentioned above so that the estimated covariance matrix can be inverted~\cite{Hartlap:2006}.

If we fix the window convolution formula to the reference case~\(\Lambda_\rf\) given by \cref{eq:window-convolution-formula-ref}, we can minim{\is}e \cref{eq:loss-function} over the amplitude offset~\(\beta\) to find the best-fitting value~\(\beta = \num{6e-3}\) for the monopole~\(\wind{B}_{000}\) and \(\beta = \num{4e-2}\) for the quadrupole~\(\wind{B}_{202}\). In both cases, this value is well within the fractional error of the cubic-box measurements (which we have used as a noisy proxy model) in this scale range.
We thus consider the offset to be consistent with zero and assume \(\beta = 0\) hereafter in this section.
For the monopole~\(\wind{B}_{000}\), the loss function~\(\chi^2_{000}\) is \num{0.08} per wavenumber bin, and for the quadrupole~\(\wind{B}_{202}\), the loss function~\(\chi^2_{202}\) is \num{0.03} per wavenumber bin.

\begin{figure}
    \centering
    {\small DESI DR1 LRG SGC \(\num{0.4} \leq z \leq \num{0.6}\)\smallskip}
    \begin{subfigure}{0.5\textwidth}
        \flushleft
        \includegraphics[width=0.975\linewidth]{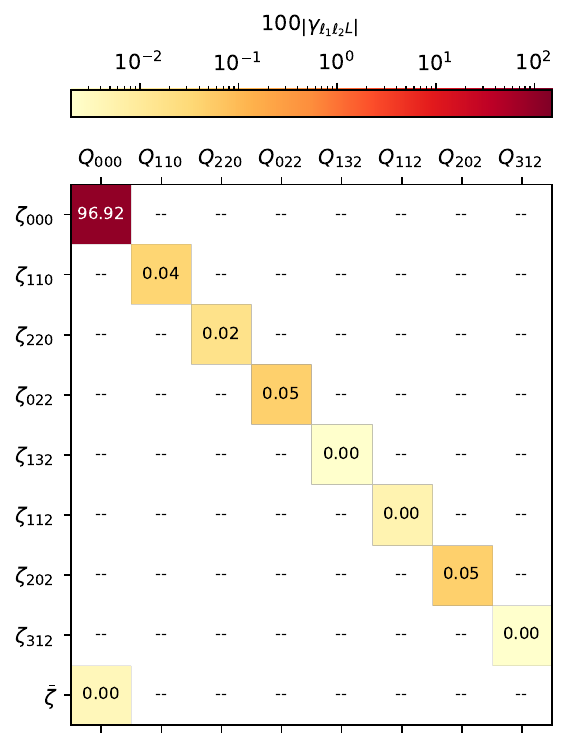}
        \caption{monopole~\(\wind{B}_{000}\) contribution}
        \label{fig:window-contribution-remove-monopole}
    \end{subfigure}%
    \begin{subfigure}{0.5\textwidth}
        \flushright
        \includegraphics[width=0.975\linewidth]{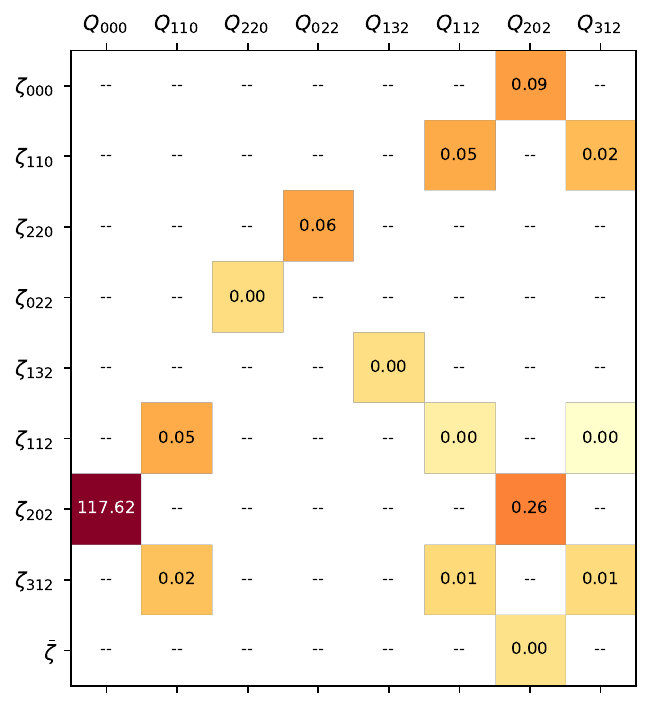}
        \caption{quadrupole~\(\wind{B}_{202}\) contribution}
        \label{fig:window-contribution-remove-quadrupole}
    \end{subfigure}
    \caption{%
        Absolute values of the weight function~\eqref{eq:weight-function}, \(100 \abs{\func{\gamma_{\ell_1 \ell_2 L}}(\Lambda)}\), for each convolution formula~\(\Lambda\) with precisely one term~\(Q_{\ell''_1 \ell''_2 L''} \zeta_{\ell'_1 \ell'_2 L'}\) removed from the reference formula~\(\Lambda_\rf\) given by \cref{eq:window-convolution-formula-ref}.
        For visual clarity, the values have been multiplied by 100 and the colour map is in a logarithmic scale.
        Sub-figure~(\subref{fig:window-contribution-remove-monopole}) shows the relative contribution of each term to the windowed bispectrum monopole~\(\wind{B}_{000}\), and sub-figure~(\subref{fig:window-contribution-remove-quadrupole}) to the windowed quadrupole~\(\wind{B}_{202}\).
    }
    \label{fig:window-contribution-remove}
\end{figure}

To evaluate the relative contribution of each term~\(Q_{\ell''_1 \ell''_2 L''} \zeta_{\ell'_1 \ell'_2 L'}\) (numerical prefactor omitted in text for brevity) in~\cref{eq:window-convolution-formula-ref}, we define the weight function
\begin{equation}
    \func{\gamma_{\ell_1 \ell_2 L}}(\Lambda) = {\frac{%
        \chi^2_{\ell_1 \ell_2 L}(\Lambda, \beta) %
        - \chi^2_{\ell_1 \ell_2 L}(\Lambda_\rf, \beta)
    }{
        \chi^2_{\ell_1 \ell_2 L}(\Lambda_0, \beta) %
        - \chi^2_{\ell_1 \ell_2 L}(\Lambda_\rf, \beta)
    }} \comma
    \label{eq:weight-function}
\end{equation}
where \(\Lambda_0\) denotes the case of a zero bispectrum model, i.e. \(\wind{B}_{\ell_1 \ell_2 L} \equiv 0\).
This weight function is akin to a sliding scale with the end-points chosen to correspond to a zero bispectrum model and the reference formula, but it should not be interpreted as a percentage contribution.%
\footnote{%
Both positive and negative terms in the window convolution formula may lead to an increased or decreased value of \(\chi^2_{\ell_1 \ell_2 L}(\Lambda, \beta)\).
}
We first consider a set of window function formulae~\(\Lambda\) each with precisely one term removed from \(\Lambda_\rf\), in which case a larger value of \(\func{\gamma_{\ell_1 \ell_2 L}}(\Lambda)\) indicates a more significant term, since its removal leads to a greater deviation in the window-convolved bispectrum model.
In \cref{fig:window-contribution-remove}, we show the absolute weight function value of the various terms in the reference window convolution formula~\eqref{eq:window-convolution-formula-ref} for the bispectrum monopole~\(\wind{B}_{000}\) and quadrupole~\(\wind{B}_{202}\) separately. 
It is clear that the latter terms in the reference formula~\eqref{eq:window-convolution-formula-ref}, e.g. those with the multipole factor~\(\zeta_{132}\) or \(\zeta_{312}\), are negligible.
The integral constraint correction is also negligible for both the monopole~\(\wind{B}_{000}\) and the quadrupole~\(\wind{B}_{202}\).

\begin{figure}
    \centering
    {\small DESI DR1 LRG SGC \(\num{0.4} \leq z \leq \num{0.6}\)\smallskip}
    \begin{subfigure}{\textwidth}
        \centering
        \includegraphics[width=0.75\linewidth,trim={0 4mm 0 0}]{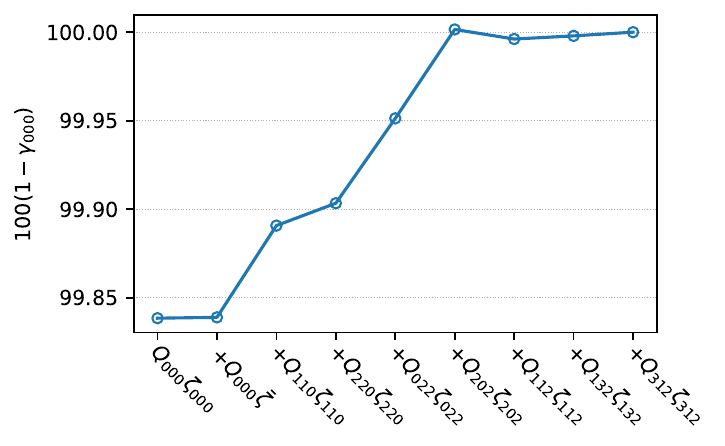}
        \caption{monopole~\(\wind{B}_{000}\) convergence}
        \label{fig:window-contribution-cascade-monopole}
    \end{subfigure}
    \smallskip
    
    \begin{subfigure}{\textwidth}
        \centering
        \includegraphics[width=0.75\linewidth]{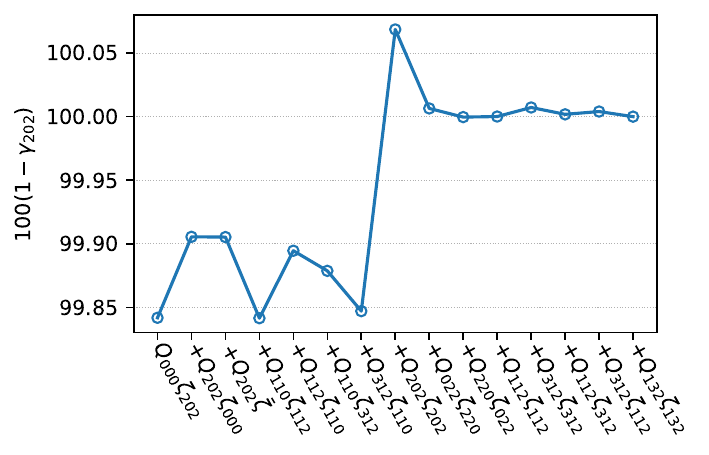}
        \caption{quadrupole~\(\wind{B}_{202}\) convergence}
        \label{fig:window-contribution-cascade-quadrupole}
    \end{subfigure}
    \caption{%
        The quantity~\(100(1 - {\gamma_{\ell_1 \ell_2 L}})\), where \(\func{\gamma_{\ell_1 \ell_2 L}}(\Lambda)\) is the weight function~\eqref{eq:weight-function}, for a series of convolution formulae~\(\Lambda\) that adds each term~\(Q_{\ell''_1 \ell''_2 L''} \zeta_{\ell'_1 \ell'_2 L'}\) cumulatively to the leading term~\(Q_{000} \zeta_{\ell_1 \ell_2 L}\).
        Sub-figure~(\subref{fig:window-contribution-cascade-monopole}) shows the convergence of the window convolution formula~\eqref{eq:window-convolution-formula-ref} for the windowed bispectrum monopole~\(\wind{B}_{000}\), and sub-figure~(\subref{fig:window-contribution-cascade-quadrupole}) for the windowed quadrupole~\(\wind{B}_{202}\).
    }
    \label{fig:window-contribution-cascade}
\end{figure}

To visual{\is}e the convergence of the window convolution series, in \cref{fig:window-contribution-cascade} we consider the quantity~\((1 - \gamma_{\ell_1 \ell_2 L})\) for a set of window function formulae~\(\Lambda\) that cumulatively adds each term in \cref{eq:window-convolution-formula-ref}, starting from the leading term, i.e. \(\wind{\zeta}_{\ell_1 \ell_2 L} = Q_{000} \zeta_{\ell_1 \ell_2 L}\).
Once sufficiently many terms have been included, the value of 
\((1 - \gamma_{\ell_1 \ell_2 L})\) should stabil{\is}e to \num{1}, as observed in \cref{fig:window-contribution-cascade}.
Based on \cref{fig:window-contribution-remove,fig:window-contribution-cascade}, we can write down a window convolution formula from a reduced set of multipoles by removing terms with~\(\abs{\gamma_{000}}, \abs{\gamma_{202}} < \num{4e-4}\),
\begin{subequations}
\begin{align}
\wind{\zeta}_{000} & =
Q_{000} \zeta_{000} + \frac{1}{3} Q_{110} \zeta_{110} + \frac{1}{5} \qty(Q_{022} \zeta_{022} + Q_{202} \zeta_{202}) \comma \\
\wind{\zeta}_{202} & =
Q_{000} \zeta_{202} + Q_{202} \zeta_{000} + \frac{1}{3} \qty(Q_{112} \zeta_{110} + Q_{110} \zeta_{112}) + \frac{1}{5} Q_{022} \zeta_{220} + \frac{2}{7} Q_{202} \zeta_{202} \fstop
\end{align}
\label{eq:window-convolution-formula-reduced}%
\end{subequations}
The loss function~\eqref{eq:loss-function} remains the same as for the full reference formula, at \num{0.08} for the monopole~\(\wind{B}_{000}\) and \num{0.03} for the quadrupole~\(\wind{B}_{202}\) per wavenumber bin.
In the next subsection, we will use this reduced window convolution formula to test the window matrix in the linear algebra formulation.

Although the validation tests in this section have focused on the LRG SGC sample in the redshift bin \(\num{0.4} \leq z \leq \num{0.6}\) in DESI DR1, we have also checked other samples such as quasars (QSO) in the North Galactic Cap~(NGC) in the redshift range \(0.8 \leq z \leq 2.1\).%
\footnote{Data products of the three-point window function for different samples will be made public at \href{https://github.com/MikeSWang/DESI-DR1-Clustering-ThreePointWindow}{{\burl{https://|github.|com/|MikeSWang/|DESI-|DR1-|Clustering-|ThreePointWindow}}} (see \hyperlink{par:data-availability}{Data Availability}).
Additional DESI DR1 data products can be generated and provided upon request.}
These results are presented in \cref{sec:window-convolution-additional-samples}.
We note that the reduced window convolution formula~\eqref{eq:window-convolution-formula-reduced} does not apply to other samples with different volumes or completeness levels, and the convergence of window function measurements and window convolution series depends on, for instance, the specific survey geometry and scales of interest.
Therefore, when deriving the window convolution formula for a different sample, the validation procedure should be repeated on a case-by-case basis, as we have done in \cref{sec:window-convolution-additional-samples}.

\subsection{Window matrix performance}

When the set of unwindowed model multipoles~\(\zeta_{\ell_1 \ell_2 L}\) included in the window convolution formula for different windowed model multipoles~\(\wind{\zeta}_{\ell_1 \ell_2 L}\) do not entirely coincide, the overall window matrix~\(\mW\) will contain sparse sub-matrices.
In such cases, it may be computationally advantageous to consider the window matrix for each required windowed multipole separately.
Without loss of generality, we compute in this subsection the window matrices for the bispectrum monopole~\(\wind{B}_{000}\) and quadrupole~\(\wind{B}_{202}\) separately, which correspond to the two reduced window convolution formulae~\eqref{eq:window-convolution-formula-reduced}.

\begin{figure}
    \centering
    {\small DESI DR1 LRG SGC \(\num{0.4} \leq z \leq \num{0.6}\)\smallskip}
    \begin{subfigure}{\textwidth}
        \centering
        \includegraphics[width=0.8\linewidth]{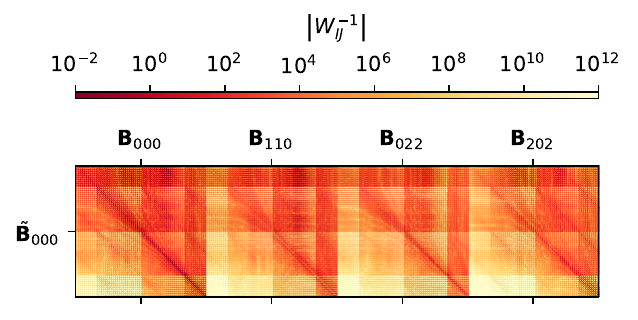}
        \caption{window matrix for monopole~\(\wind{B}_{000}\)}
        \label{fig:window-matrix-reduced-monopole}
    \end{subfigure}
    \smallskip
    
    \begin{subfigure}{\textwidth}
        \centering
        \includegraphics[width=0.925\linewidth]{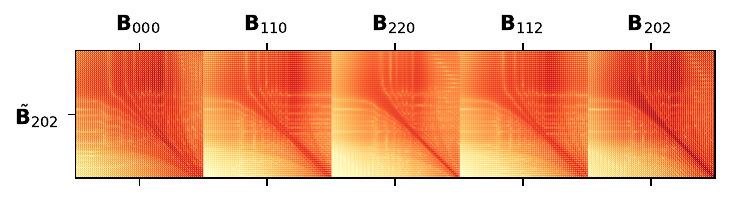}
        \caption{window matrix for quadrupole~\(\wind{B}_{202}\)}
        \label{fig:window-matrix-reduced-quadrupole}
    \end{subfigure}
    \caption{%
        Inverse absolute value of the window matrix elements, \(\qty(\mW)_{IJ}^{-1}\), for the windowed bispectrum monopole~\(\wind{B}_{000}\) (sub-figure~\subref{fig:window-matrix-reduced-monopole}) and quadrupole~\(\wind{B}_{202}\) (sub-figure~\subref{fig:window-matrix-reduced-quadrupole}) according to the reduced window convolution formula~\eqref{eq:window-convolution-formula-reduced}.
        The column blocks of the matrix corresponds to the input unwindowed bispectrum multipole vectors~\(\vB_{\ell_1 \ell_2 L}\), and the rows corresponds to the output windowed bispectrum multipole vectors~\(\vwB_{{\ell_1 \ell_2 L}}\).
    }
    \label{fig:window-matrix-reduced}
\end{figure}

The input bispectrum multipoles~\(\func{B_{\ell_1 \ell_2 L}}(k_1, k_2)\) are as before the proxy model obtained from the mean of the cubic-box measurements, which have been resampled from \num{50} to \(\num{64} = 2^6\) logarithmic bins in \(k_{1,2}\).%
\footnote{This is the nearest power of two above, which is a more natural sample number for the FFTLog algorithm.}
We fix the ordering of the multipoles as they appear in \cref{eq:window-convolution-formula-reduced} to form the vector{\is}ed bispectrum model vector~\(\vB\).
To generate the window matrix~\(\mW\), we feed a basis of unit vectors as~\(\vB\) into the full window convolution recipe as described in \cref{sec-sub:convolution-matrix-formulation}.
This results in window matrices of dimensions~\((64^2, 4 \times 64^2)\) for the monopole~\(\wind{B}_{000}\) and \((64^2, 5 \times 64^2)\) for the quadrupole~\(\wind{B}_{202}\), since the input bispectrum multipoles are two-dimensional before vector{\is}ation and there are respectively four and five unwindowed model multipoles required by them in \cref{eq:window-convolution-formula-reduced}.
Although the sub-matrices of \(\mW\) here are square with dimensions \((64^2, 64^2)\) as the numbers of input and output FFTLog sample points are the same, generally they can be non-square with more columns (input sample points) than rows (output sample points); to compare with the windowed cut-sky bispectrum data vector with fewer wavenumber bins, a simple interpolation step can be performed after convolution.
In \cref{fig:window-matrix-reduced}, we show the structure of the window matrices~\(\mW\) for the windowed bispectrum monopole~\(\wind{B}_{000}\) and quadrupole~\(\wind{B}_{202}\).

Whether we compute the windowed bispectrum multipole from the window matrix multiplication~\(\vwB = \mW \vB\) or following the step-by-step recipe in \cref{sec-sub:convolution-matrix-formulation}, the results are found to agree within floating-point arithmetic precision.
However, the full convolution pipeline without using the window matrix takes order \SI{e-1}{\second} on a single processor core, whereas the window matrix multiplication takes order \SI{e-2}{\second}.
While the initial computation of the window matrix can be relatively expensive, it can be easily parallel{\is}ed and only needs to be done once, as long as the wavenumber sample points remain the same; by contrast, window convolution needs to be performed repeatedly in a cosmological inference analysis.
Finally, in \cref{fig:matrix-multiplication-time} in \cref{sec:window-matrix-multiplication-time}, we show the matrix multiplication time for a range of matrix dimensions similar to the window matrix considered in this section.
We note that the execution time is all below \SI{e-1}{\second}, and thus the linear algebra formulation of window convolution will be performant and well-suited for cosmological likelihood analyses.

\section{Conclusion}
\label{sec:conclusion}

Window convolution for the galaxy clustering bispectrum has long been a challenging issue: its formulation is usually analytically complex, and the window function itself is computationally expensive to estimate.
The problem stems from the so-called `curse of dimensionality', as the 3PCF or bispectrum in Fourier space has more degrees of freedom than the two-point correction function~(2PCF) and the power spectrum.
With the advent of Stage~IV spectroscopic surveys including DESI and \textit{Euclid}, non-Gaussian clustering statistics have become a frontier in driving the exploration of cosmological models; as such, the survey window effect which modulates both the amplitude and shape of the three-point clustering statistics must be properly accounted for through forward modelling.

We have built on the TripoSH formalism developed in refs.~\cite{Sugiyama:2019,Sugiyama:2020} to address some of these difficulties.
In this decomposition, the bispectrum multipoles in redshift space are functions of two wavenumbers instead of a full triangle configuration, thus providing a natural form of data reduction and avoiding the triangle binning issue.
At the same time, the window convolution series derived in this formalism are structurally similar to that of the 2PCF/power spectrum \cite{Wilson:2016,Beutler:2016,Beutler:2021}, and require window function multipoles that can be conveniently estimated using FFTs in the same way as for the 3PCF.
In particular, by util{\is}ing the linear algebra nature of the window convolution operations, we have proposed a formulation that encapsulates the full procedure as a window matrix multiplication.

In this work, we have focused on tracer samples in the DESI DR1 data release, and employed a suite of unwindowed cubic-box snapshots and windowed cut-sky catalogues with the same underlying clustering signal, to validate the window convolution pipeline for the bispectrum.
We have checked the convergence of the window function measurements with respect to FFT sampling and the number density of the random catalogue, and highlighted some of the technicalities that differ from the case of the two-point window function.
We have also systematically verified the convergence of the window convolution series truncated at a finite but large number of terms, which has been either ignored or performed in configuration space for 3PCF in previous works~\cite{Sugiyama:2019,Sugiyama:2020}.
The window matrix formulation is found to provide the same results as the full convolution procedure, but with a computation time that is an order of magnitude lower once it has been generated.

The results in this work offer a replicable window convolution pipeline to implement for generic survey configurations, with all the computational tools publicly available as part of the \codename{Triumvirate} code (see \cref{fn:Triumvirate} or Data Availability for the link).
However, we caution that the convergence of the window function measurements and the window convolution series depends on the specific survey geometry and scales of interest, and should be verified in each case as detailed in \cref{sec:validation}.

Finally, some topics remain to be explored, such as the performance of alternative algorithms for the double spherical Bessel transform and the scale-dependent integral constraint, including the radial integral constraint. We leave these issues for future studies.

\acknowledgments

We would like to thank Hector Gil-Mar\'in and {Zachary Slepian for their helpful feedback on the manuscript.
We would also like to thank Ashley Ross and Arnaud de Mattia for help with the mock catalogues, and Naonori Sugiyama for useful discussions.
This project has received funding from the European Research Council under the European Union’s Horizon 2020 research and innovation programme (grant agreement 853291).
FB acknowledges the support of the Royal Society through the University Research Fellowship.

This research used data obtained with the Dark Energy Spectroscopic Instrument~(DESI).
DESI construction and operations is managed by the Lawrence Berkeley National Laboratory.
This material is based upon work supported by the United States Department of Energy~(DOE), Office of Science, Office of High-Energy Physics, under Contract No. DE-AC02-05CH11231, and by the National Energy Research Scientific Computing Center, a DOE Office of Science User Facility under the same contract.
Additional support for DESI was provided by the United States National Science Foundation~(NSF), Division of Astronomical Sciences under Contract No. AST-0950945 to the NSF’s National Optical-Infrared Astronomy Research Laboratory; the Science and Technologies Facilities Council~(STFC) of the United Kingdom; the Gordon and Betty Moore Foundation; the Heising-Simons Foundation; the French Alternative Energies and Atomic Energy Commission~(CEA); the National Council of Science and Technology of Mexico~(CONACYT); the Ministry of Science and Innovation of Spain~(MICINN), and
by the DESI Member Institutions: \url{www.desi.lbl.gov/collaborating-institutions}.
The DESI Collaboration is honoured to be permitted to conduct scientific research on Iolkam Du’ag (Kitt Peak), a mountain with particular significance to the Tohono O’odham Nation.
Any opinions, findings, and conclusions or recommendations expressed in this material are those of the author(s) and do not necessarily reflect the views of the NSF, the DOE, or any of the listed funding agencies.

This work used the DiRAC Data Intensive service at the University of Leicester~(DIaL3), managed by the University of Leicester Research Computing Service on behalf of the STFC DiRAC HPC Facility (\url{dirac.ac.uk}).
The DiRAC service at Leicester was funded by the Department for Business, Energy \& Industrial Strategy (BEIS), UK Research and Innovation (UKRI) and STFC capital funding and operations grants. 
DiRAC is part of the UKRI Digital Research Infrastructure.

\hypertarget{par:data-availability}{}\paragraph*{Data Availability.} The data underlying this article will be made public along with DESI DR1 (see \href{https://data.desi.lbl.gov/doc}{\burl{data.|desi.|lbl.|gov/|doc}} for details) as part of the DESI Data Management Plan.

The source code used to process and generate the relevant data in this article is made freely available under the GPLv3+ licence in the GitHub repository at \href{https://github.com/MikeSWang/Triumvirate}{\burl{github.|com/|MikeSWang/|Triumvirate}}.

\appendix

\section{Window convolution tests of additional samples}
\label{sec:window-convolution-additional-samples}

We have additionally studied the DESI DR1 QSO sample which has a single redshift bin \(0.8 \leq z \leq 2.1\) in both NGC and SGC, and is lower in number density and higher in redshift compared to the LRG sample studied in \cref{sec:validation}.
For the QSO NGC sample which has a much larger comoving volume of \SI{15.7}{\cubic\per\h\cubic\giga\parsec} (and an effective volume of \SI{1.0}{\cubic\per\h\cubic\giga\parsec} only due to low number density)~\cite{DESI2024.II.KP3,DESI2024.V.KP5}, we measure the window function multipoles~\({Q_{\ell_1 \ell_2 L}}\) using a random catalogue \(\alpha^{-1} \approx 127\) times denser than the data catalogue with a mesh grid of minimum dimension~\(L = \SI{8000}{\lunit}\) and cell size~\(\Delta = \SI{7.8}{\lunit}\), which is limited by the computational memory available.%
\footnote{Although \cref{fig:window-convergence-fftsamp-cellsize} shows that this relatively large cell size leads to non-negligible deviations in the measured window function multipoles, these are subdominant to the larger sample variance of the QSO sample and do not noticeably impact the final window convolution results seen in \cref{fig:window-convolution-reduced-add-QSO-NGC}.}
Both data and random mock catalogues have been generated from the ATMLs which simulate fibre assignment in DESI observations.

Starting from the same reference formula~\eqref{eq:window-convolution-formula-ref} and the weight function~\eqref{eq:weight-function} in \cref{sec-sub:validation-series}, we derive the convergent window convolution series based on the criterion~\(\abs{\gamma_{\ell_1 \ell_2 L}} \geq \num{5e-4}\),
\begin{subequations}
\begin{align}
\wind{\zeta}_{000} & =
Q_{000} \zeta_{000} + \frac{1}{3} Q_{110} \zeta_{110} + \frac{1}{5} Q_{220} \zeta_{220} - Q_{000} \zetabar \comma \\
\wind{\zeta}_{202} & =
Q_{000} \zeta_{202} + Q_{202} \zeta_{000} + \frac{1}{3} \qty(Q_{112} \zeta_{110} + Q_{110} \zeta_{312} + Q_{312} \zeta_{110}) \nonumber \\
& \qquad + \frac{1}{5} Q_{022} \zeta_{220} + \frac{2}{7} Q_{202} \zeta_{202} - Q_{202} \zetabar \fstop
\end{align}
\end{subequations}
In \cref{fig:window-convolution-reduced-add-QSO-NGC}, we compare the window-convolved bispectrum proxy model~\(\wind{B}_{\ell_1 \ell_2 L}\) with the windowed cut-sky measurements and the unwindowed cubic-box measurements in the wavenumber range~\((0, 0.12]~\si{\wunit}\).
\begin{figure}
    \centering
    {\small DESI DR1 QSO NGC \(\num{0.8} \leq z \leq \num{2.1}\)\smallskip}
    \includegraphics[width=\linewidth]{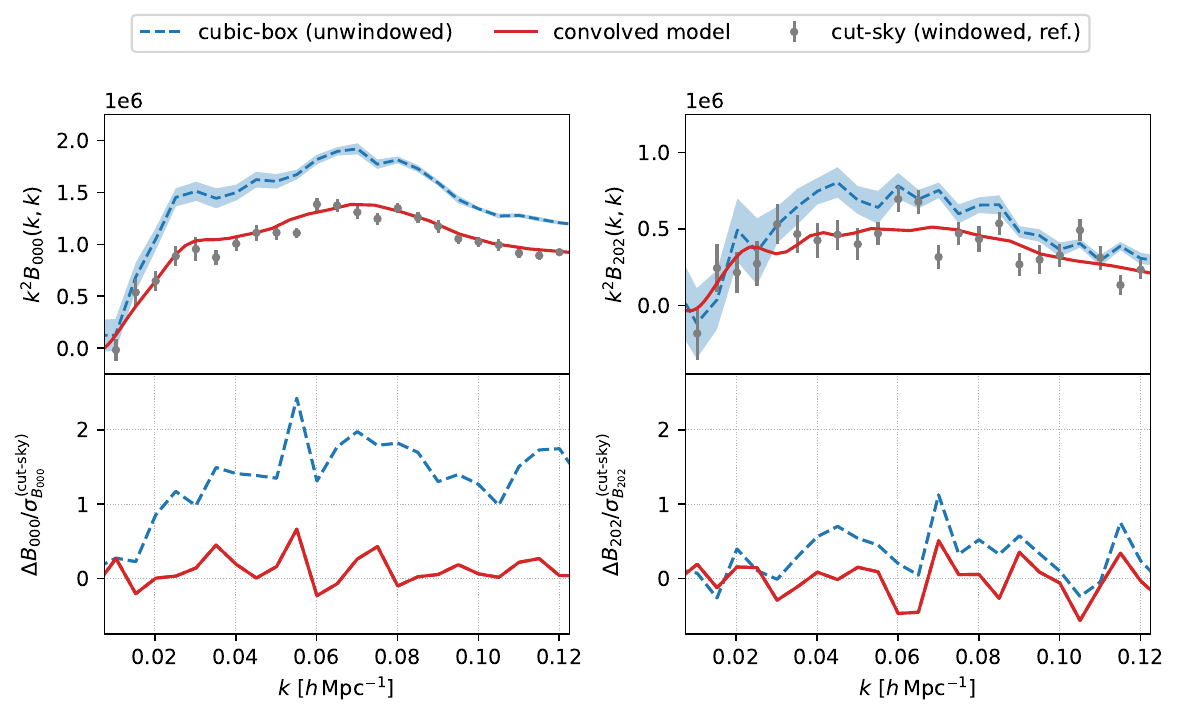}
    \caption{%
        The same as \cref{fig:window-convolution-full} but for the DESI DR1 QSO NGC sample in the redshift bin \(\num{0.8} \leq z \leq \num{2.1}\).
    }
    \label{fig:window-convolution-reduced-add-QSO-NGC}
\end{figure}
As before, the effect of the window function is significant especially for the bispectrum monopole~\(\wind{B}_{000}\), but to a lesser extent than the LRG SGC sample as the larger survey volume reduces the impact of the window function in this scale range.
The window-convolved model matches the amplitude and shape of the cut-sky measurements well, with the loss function~\(\chi^2_{\ell_1 \ell_2 L}\) (eq.~\ref{eq:loss-function}) per wavenumber bin valued at \num{0.03} for the monopole~\(\wind{B}_{000}\) and \num{0.12} for the quadrupole~\(\wind{B}_{202}\).
If we allow for a relative amplitude offset~\(\beta\) and minim{\is}e \cref{eq:loss-function} over it, we find \(\beta\) consistent with zero offset for both the monopole~\(\wind{B}_{000}\) and quadrupole~\(\wind{B}_{202}\) within the fractional error of the cubic-box measurements used as the noisy proxy model.

However, when we test the window convolution pipeline on complete LRG mock catalogues without fibre assignment (in the SGC redshift bin \(\num{0.4} \leq z \leq \num{0.6}\)), we find there is a slightly larger but still small amplitude offset.
The convergent window convolution series satisfying \(\abs{\gamma_{\ell_1 \ell_2 L}} \geq \num{5e-4}\) are given by
\begin{subequations}
\begin{align}
\wind{\zeta}_{000} & =
Q_{000} \zeta_{000} + \frac{1}{3} Q_{110} \zeta_{110} + \frac{1}{5} \qty(Q_{022} \zeta_{022} + Q_{220} \zeta_{220}) - Q_{000} \zetabar \comma \\
\wind{\zeta}_{202} & =
Q_{000} \zeta_{202} + Q_{202} \zeta_{000} + \frac{1}{3} \qty(Q_{112} \zeta_{110} + Q_{110} \zeta_{112}) + \frac{2}{7} Q_{202} \zeta_{202} - Q_{202} \zetabar \fstop
\end{align}
\end{subequations}
When the loss function~\(\chi^2_{\ell_1 \ell_2 L}\) is minim{\is}ed over \(\beta\), we find \(\beta = \num{3e-2}\) for the monopole~\(\wind{B}_{000}\) and \(\beta = \num{4e-2}\) for the quadrupole~\(\wind{B}_{202}\), which remain within the fractional error of the measurements.
In \cref{fig:window-convolution-reduced-add-LRG-SGC-complete}, we make the same comparison as before between windowed and unwindowed bispectrum multipoles including the window-convolved model rescaled by \((1 + \beta)\) to account for the amplitude offset.
\begin{figure}
    \centering
    {\small DESI DR1 LRG SGC \(\num{0.4} \leq z \leq \num{0.6}\) (complete mock catalogues)\smallskip}
    \includegraphics[width=\linewidth]{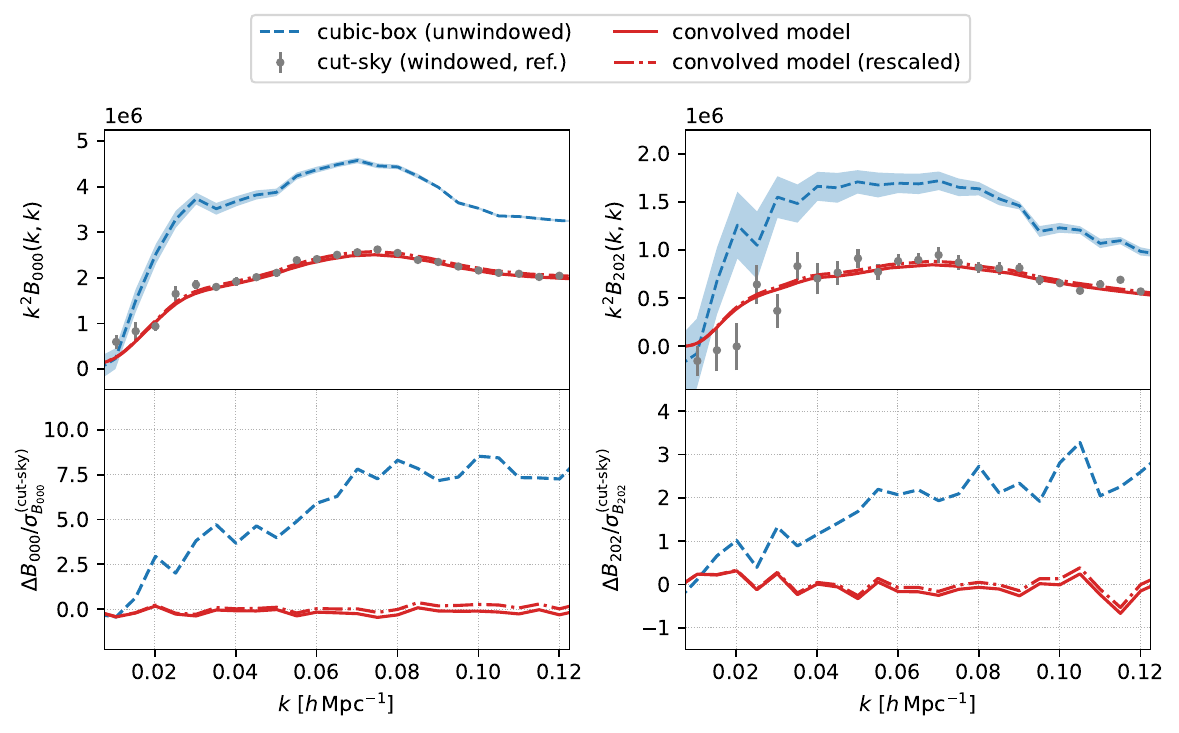}
    \caption{%
        The same as \cref{fig:window-convolution-full} but for the complete mock catalogues.
        In addition, the window-convolved proxy model rescaled by \((1 + \beta)\) is shown where \(\beta = \num{3e-2}\) and \num{4e-2} account for a small amplitude offset in the monopole and quadrupole respectively.
    }
    \label{fig:window-convolution-reduced-add-LRG-SGC-complete}
\end{figure}
With this amplitude rescaling, the loss function value per wavenumber bin reduces from \num{0.09} to \num{0.06} for the monopole~\(\wind{B}_{000}\) and \num{0.12} to \num{0.11} for the quadrupole~\(\wind{B}_{202}\).
We have checked that the optimal value of \(\beta\) does not change significantly with the measurement parameters of the window function multipoles such as the mesh grid physical size, cell resolution and the random catalogue number density, or when the full reference formula is used.

We have repeated the same checks with the complete QSO mock catalogues in both NGC and SGC in the redshift bin \(\num{0.8} \leq z \leq \num{2.1}\), but do not find any significant amplitude offset (see \cref{{fig:window-convolution-reduced-add-QSO-NGC-complete}} for the NGC case).
\begin{figure}
    \centering
    {\small DESI DR1 QSO NGC \(\num{0.8} \leq z \leq \num{2.1}\) (complete mock catalogues)\smallskip}
    \includegraphics[width=\linewidth]{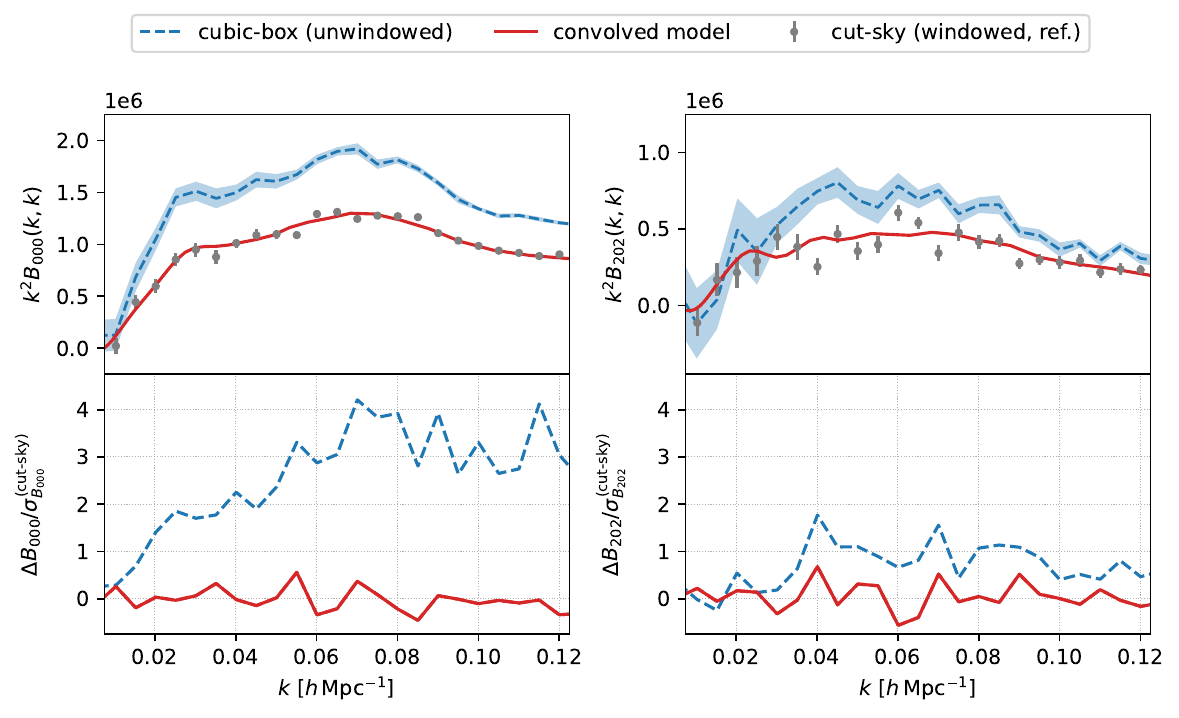}
    \caption{%
        The same as \cref{fig:window-convolution-reduced-add-QSO-NGC} but for the complete mock catalogues.
    }
    \label{fig:window-convolution-reduced-add-QSO-NGC-complete}
\end{figure}
One possible explanation is that there may be a small amount of unknown systematic effects specific to the complete LRG mock catalogues; in particular, we note that the AMTL mock catalogues are not derived from the complete ones, but instead both have been separately generated from potential-assignment targets~\cite{Lasker:2024,DESI2024.II.KP3}. A second possibility is the imperfect shot noise subtraction~\cite{Sugiyama:2019} in the proxy `models' based on mock measurements used in the validation tests here; this, however, would not be the case in an actual cosmological analysis which forward models theoretical predictions. With that being said, we caution that these interpretations lie outside the scope of the window convolution treatment itself and should be separately examined. The relative amplitude offset parameter~\(\beta\), as well as being a diagnostic tool in conjunction with the loss function~\(\chi^2_{\ell_1 \ell_2 L}\), also serves as a useful nuisance parameter to absorb unknown systematic residuals after window convolution.
In the cosmological analysis of DESI DR1 samples, it is the AMTL not complete mock catalogues that will be used to validate the full pipeline as they resemble the actual survey data more closely; therefore, the validation of our window convolution pipeline with the AMTL mock catalogues in this work is more important.

\section{Window matrix multiplication computation time}
\label{sec:window-matrix-multiplication-time}

We consider a range of window matrix dimensions~\((Nd, 4Nd)\), where \(d = 1024\) is the base dimension and \(N \in \qty{2, 4, 8, 12, 16}\) is an integer multiplier.
For instance, the window matrix~\(\mW\) for the windowed bispectrum monopole~\(\wind{B}_{000}\) used in \cref{sec-sub:validation-series} has \(N = 4\).
In \cref{fig:matrix-multiplication-time}, we plot the measured processor time of matrix multiplication for random matrices generated with these dimensions and compare three matrix multiplication methods available in the public \textsc{NumPy} Python package~\cite{NumPy:2020}.
These multiplication methods perform similarly well, with an execution time all below \SI{e-1}{\second}; therefore, the window matrix multiplication step will not be the dominant operation in a typical cosmological likelihood analysis.
\begin{figure}
    \centering
    \includegraphics[width=0.75\linewidth]{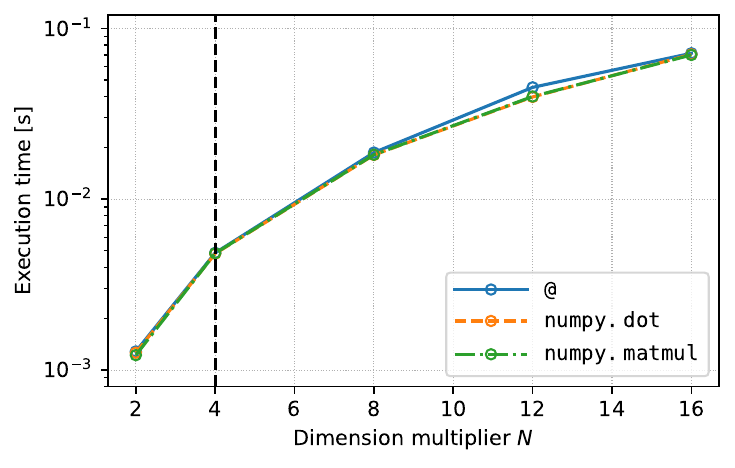}
    \caption{%
        Matrix multiplication time for a range of possible dimensions~\((Nd, 4Nd)\) for the window matrix~\(\mW\), measured for three matrix multiplication methods available in the public \textsc{NumPy} Python package.
        Here \(d = 1024\) is the base dimension considered, and \(N\) is an integer multiplier.
        The dashed vertical line at \(N = 4\) marks the (row) dimension of the window matrix~\(\mW\) used for the windowed bispectrum monopole~\(\wind{B}_{000}\) in \cref{sec-sub:validation-series}.
    }
    \label{fig:matrix-multiplication-time}
\end{figure}

\bibliographystyle{JHEP}
\bibliography{bibliography,desipub}

\end{document}